\documentclass[manuscript,screen]{acmart}

\usepackage{bm}
\usepackage{booktabs}
\usepackage{enumitem}
\usepackage{multirow}
\usepackage{algorithm}
\usepackage[noEnd,commentColor=blue]{algpseudocodex}
\usepackage{tablefootnote}

\renewcommand{\paragraph}[1]{\noindent\textbf{#1.}}

\DeclareMathOperator*{\argmax}{arg\,max}

\copyrightyear{2024}
\acmYear{2024}
\setcopyright{rightsretained}

\begin{document}

\title[DPSW-Sketch: A Differentially Private Sketch Framework for Frequency Estimation over Sliding Windows]{DPSW-Sketch: A Differentially Private Sketch Framework for Frequency Estimation over Sliding Windows (Technical Report)}

\author{Yiping Wang}
\orcid{0009-0009-7526-985X}
\affiliation{%
  \institution{East China Normal University}
  \city{Shanghai}
  \country{China}}
\email{yipwang@stu.ecnu.edu.cn}

\author{Yanhao Wang}\authornote{Corresponding author.}
\orcid{0000-0002-7661-3917}
\affiliation{%
  \institution{East China Normal University}
  \city{Shanghai}
  \country{China}}
\email{yhwang@dase.ecnu.edu.cn}

\author{Cen Chen}\authornote{Cen Chen is also affiliated with The State Key Laboratory of Blockchain and Data Security, Zhejiang University, Hangzhou, China.}
\orcid{0000-0003-0325-1705}
\affiliation{%
  \institution{East China Normal University}
  \city{Shanghai}
  \country{China}}
\email{cenchen@dase.ecnu.edu.cn}

\begin{abstract}
The sliding window model of computation captures scenarios in which data are continually arriving in the form of a stream, and only the most recent $w$ items are used for analysis. In this setting, an algorithm needs to accurately track some desired statistics over the sliding window using a small space. When data streams contain sensitive information about individuals, the algorithm is also urgently needed to provide a provable guarantee of privacy. In this paper, we focus on the two fundamental problems of privately (1) estimating the frequency of an arbitrary item and (2) identifying the most frequent items (i.e., \emph{heavy hitters}), in the sliding window model. We propose \textsc{DPSW-Sketch}, a sliding window framework based on the count-min sketch that not only satisfies differential privacy over the stream but also approximates the results for frequency and heavy-hitter queries within bounded errors in sublinear time and space w.r.t.~$w$. Extensive experiments on five real-world and synthetic datasets show that \textsc{DPSW-Sketch} provides significantly better utility-privacy trade-offs than state-of-the-art methods.
\end{abstract}




\maketitle

\section{Introduction}
\label{sec-intro}

Many real-world applications, including Internet of Things (IoT), location-based services, and public health surveillance, generate large volumes of time-evolving data.
These data are typically collected continuously in the form of \emph{streams}.
Therefore, the design and analysis of streaming algorithms \cite{Muthukrishnan05}, which primarily aim to provide accurate estimations over the stream in real time using a small space, has attracted much attention in the literature.

As data streams often contain sensitive information about individuals, such as browsing history, GPS coordinates, and health status, releasing the estimates computed on them without adequate protection can cause harmful privacy leakage.
To address this issue, differential privacy \cite{DworkMNS06} (DP), the de facto standard for privacy preservation, has been widely adopted in streaming analytics \cite{DworkNPR10, ChanSS10}.
Informally, a randomized algorithm provides (event-level) DP if the output distribution of the algorithm is approximately the same when executed on a stream and any of its neighbors that differ from it by one item.
This condition prevents an attacker with access to the output from inferring the existence of any individual item.

Furthermore, in practice, it is often essential to restrict the computation to recent data.
For example, Apple's DP team claims to retain the collected data for a maximum of three months \cite{Upadhyay19}, and Google's privacy policy states that browser data can be stored for up to nine months \cite{GooglePrivacy}.
Such recency requirements are captured by the sliding window model \cite{DatarGIM02, BravermanO07}, where only the latest $w$ items in the stream are used for computation.
In such scenarios, an algorithm should maintain and release real-time statistics continuously over the sliding window with differential privacy.
Especially, we focus on two fundamental data statistics with numerous applications, including anomaly detection \cite{ZhongYLTYC21}, change detection \cite{HaugBZK22}, and autocomplete suggestions \cite{ZhuKMSL20}, namely, \emph{frequency}, i.e., the number of occurrences of any given item, and \emph{heavy hitters}, i.e., the set of items that appear at least a given number of times.

For frequency estimation and heavy-hitter detection problems, sketches, which are compact probabilistic data structures to summarize streams \cite{CormodeGHJ12}, have been widely employed for their high accuracy and efficiency.
However, despite several attempts to extend the sketches to work in the sliding window model while guaranteeing differential privacy \cite{ChanLSX12, Upadhyay19, abs-2302-11081}, these approaches have two main limitations that make them incapable of meeting the need for real-time stream analysis.
First, their theoretical analyses are specific to heavy hitters and cannot provide any error bound for frequency estimation on the remaining items.
Second, even for heavy hitters, these methods still fail to obtain accurate query results within reasonable privacy budgets and sketch sizes in practice.

\smallskip
\paragraph{Our Contributions}
In this work, we propose a novel sketch framework with event-level differential privacy, called \textsc{DPSW-Sketch}, for frequency estimation in the sliding window model.
Specifically, \textsc{DPSW-Sketch} divides the stream into disjoint substreams of equal length.
Within each substream, it creates a series of checkpoints according to the idea of \emph{smooth histograms} \cite{BravermanO07}.
Then, it constructs a Private Count-Min Sketch \cite{ZhaoQRAAW22} (PCMS) for each checkpoint, which corresponds to all items either from the beginning of the substream to the checkpoint or from the checkpoint to the end of the substream.
As such, it does not need to store any item after the item is processed, and it can maintain each PCMS incrementally without handling implicit item deletion over the sliding window.
Furthermore, we design an efficient scheme to allocate the privacy budget across different PCMSs in each substream such that \textsc{DPSW-Sketch} provides $(\varepsilon, \delta)$-differential privacy for any given $\varepsilon > 0$ and $\delta \in (0, 1)$ throughout the stream, owing to the reduction from zero-Concentrated Differential Privacy (zCDP) \cite{BunS16} to DP and the adaptive composition of zCDP.
For any frequency or heavy-hitter query on the sliding window $W_t$ at any time $t$, it selects one PCMS per substream so that the concatenation of their corresponding items is closest to $W_t$ and combines the item frequencies obtained from all selected PCMSs to compute the result.
Our theoretical analysis shows that \textsc{DPSW-Sketch} can estimate the frequency of an arbitrary item in $W_t$ within a bounded additive error.
This approximation can be naturally extended to the problem of heavy-hitter identification.
Meanwhile, \textsc{DPSW-Sketch} has sublinear time and space complexities w.r.t.~the window size $w$.
Finally, extensive experimental findings on five real-world and synthetic datasets confirm the superior performance of \textsc{DPSW-Sketch} compared to several state-of-the-art differentially private sliding-window sketches for frequency and heavy-hitter queries.
In summary, the main contributions of this paper include:
\begin{itemize}
  \item We present the \textsc{DPSW-Sketch} framework and thoroughly analyze its approximation guarantees for frequency and heavy-hitter queries as well as time and space complexities. In particular, we indicate that \textsc{DPSW-Sketch} achieves improved theoretical results over existing methods \cite{ChanLSX12, Upadhyay19, abs-2302-11081} with lower error bounds and/or complexities.
  \item We demonstrate that \textsc{DPSW-Sketch} provides much more accurate results for frequency and heavy-hitter queries using a smaller space than the sketches proposed in \cite{ChanLSX12, Upadhyay19, abs-2302-11081} under the same privacy budget.
\end{itemize}

\paragraph{Reproducibility}
Our code and data are publicly available at \url{https://github.com/wypsz/DP-Sliding-Window}.

\section{Related Work}
\label{sec-lit}

Sketching techniques \cite{CormodeGHJ12} have attracted long-standing attention in the literature.
As for frequency estimation and heavy-hitter detection problems, classic sketching methods include Count-Min Sketch (CMS) \cite{CormodeM05}, CountSketch \cite{CharikarCF02}, CU sketch \cite{EstanV02}, Misra-Gries (MG) sketch \cite{MisraG82}, etc.
On the basis of these sketches, numerous non-private frameworks were proposed for frequency and heavy-hitter queries over sliding windows, e.g., \cite{ArasuM04, PapapetrouGD12, RivettiBM15, Ben-BasatEFK16, GouHZWLYW020, 0001JDZCH0U023}.
Unfortunately, the above frameworks do not provide any privacy guarantee and can potentially leak sensitive user data.

Differentially private sketches for frequency and heavy-hitter queries were proposed in \cite{MirMNW11, MelisDC16, YildirimKAE20, HuangQYC22, ZhaoQRAAW22, PaghT22, LebedaT23, FichtenbergerHU23, AnderssonP23}.
These sketches can provide provable guarantees of both privacy and accuracy while not reducing efficiency.
However, they cannot handle implicit deletions of items and thus cannot work in the sliding window model.
Among the existing literature, the works most related to ours are \cite{ChanLSX12, Upadhyay19, abs-2302-11081}, which proposed sketching methods for heavy-hitter identification with event-level differential privacy in the sliding window model.
\citet{ChanLSX12} first proposed a differentially private algorithm based on MG sketches for heavy-hitter identification in the (distributed) sliding window model.
\citet{Upadhyay19} designed a sliding-window framework to privately identify heavy hitters and estimate their frequencies based on CountSketch and CMS.
\citet{abs-2302-11081} proposed an improved method for private heavy-hitter queries by combining MG sketches with hierarchical counters.
Compared to \cite{ChanLSX12, Upadhyay19, abs-2302-11081}, our work achieves better theoretical results by (1) providing approximations for the frequencies of all items, instead of only those of heavy hitters, and (2) reducing the error bound for heavy hitters and complexity (see Section~\ref{subsec-theory} for a detailed comparison).
Our proposed method also demonstrates substantial improvements in practical performance compared to the methods proposed in \cite{ChanLSX12, Upadhyay19, abs-2302-11081}, as will be shown in Section~\ref{sec-experiments}.

Differentially private sketches for other stream statistics, such as quantiles \cite{PerrierAK19, GillenwaterJK21, KaplanSS22, EliezerMZ22}, moments \cite{WangPS22, EpastoMMMVZ23}, and cardinality \cite{Smith0T20, HehirTC23, Ghazi0NM23}, have also received significant attention.
The sketches employed in the above problems differ from those in frequency estimation and thus are not comparable to our framework in this paper.

\section{Preliminaries}
\label{sec-definition}

For a positive integer $n$, we use $[n]$ to denote $[1, \dots, n]$.
A data stream is an (ordered) sequence of $n$ items $S = [s_1, s_2, \dots, s_n]$, where $s_i$ is the $i$-th item and is drawn from a domain $\mathcal{E}$ of size $m$ indexed by $[m]$.
Then, $S_t = [s_1, \dots, s_t]$ denotes the prefix of $S$ until the arrival of $s_t$ at any time $t \leq n$.
A \emph{count-based sliding window} $W_t$ of length $w \in \mathbb{Z}^{+}$ at time $t$ is the union of the latest $w$ items in $S_t$, i.e., $W_t = \{s_i \in S_t : \max(t - w + 1, 1) \leq i \leq t\}$.
Note that our framework can be applied to other types of sliding windows, e.g., time-based sliding windows that contain all the items in the last $w$ time units.
For ease of presentation, we focus on count-based sliding windows in this work. 
We consider two fundamental statistics over sliding windows: \emph{frequency} and \emph{heavy hitter}.
Specifically, the frequency $f_{e, t}$ of an item $e \in \mathcal{E}$ in $W_t$ is the number of occurrences of $e$ in $W_t$, i.e., $f_{e, t} = \left|\{ s_i \in W_t : s_i = e \}\right|$.
Items with high frequencies are often called \emph{heavy hitters}.
Given a threshold $\gamma \in (0, 1)$, an item is a $\gamma$-heavy hitter in $W_t$ if its frequency in $W_t$ is at least $\gamma w$.
Formally, the set of $\gamma$-heavy hitters in $W_t$ is defined as $\mathcal{H}_{\gamma, t} = \{e \in \mathcal{E} : f_{e, t} \geq \gamma w\}$.

\paragraph{Smooth Histogram}
The smooth histogram \cite{BravermanO07} is a sublinear data structure to transform an insert-only streaming algorithm to work in the sliding window model, based on the concept of smooth functions.
Let $A$, $B$, and $C$ be the substreams of $S$ such that $B$ is a suffix of $A$ and $C$ is adjacent to $A$ and $B$.
A nonnegative function $g$ defined on any substream of $S$ is $(\alpha_1, \alpha_2)$-smooth if $(1 - \alpha_2) g(A) \leq g(B)$ implies $(1 - \alpha_1) g(A \cup C) \leq g(B \cup C)$ for two parameters $0 < \alpha_2 \leq \alpha_1 < 1$ and any above-defined $A$, $B$, and $C$.
Our analysis in Section~\ref{sec-method} will use the following theorem on the approximation and complexity of the smooth histogram.
\begin{theorem}\label{thm:smooth_histogram}
    Given an $(\alpha_1, \alpha_2)$-smooth function $g$ for two parameters $0 < \alpha_2 \leq \alpha_1 < 1$, suppose that there is an insert-only streaming algorithm $\mathcal{A}$ that produces a $\gamma$-approximation of $g$ using space $\mathcal{S}$ and update time $\mathcal{T}$. Then, there exists a sliding-window algorithm that outputs a $(\gamma + \alpha_1)$-approximation of $g$ using space $O(\frac{\log w}{\alpha_2} ( \mathcal{S} + \log w))$ and update time $O(\frac{1}{\alpha_2} \mathcal{T} \log w)$.
\end{theorem}

\paragraph{Privacy Definition}
In this work, we adopt the definition of \emph{event-level differential privacy} \cite{DworkMNS06, DworkNPR10}, where a server receives the stream $S$ and builds a data structure to answer specific queries on $S$ subject to the constraint that any single event in $S$ is indistinguishable from query results.
Formally, two streams $S = [s_1, \dots, s_n]$ and $S' = [s'_1, \dots, s'_n]$ are \emph{neighboring} if they differ only in the $i$-th item, i.e., $s_i \neq s'_i$ for some $i \in [n]$ and $s_j = s'_j, \, \forall j \neq i$.
For any parameters $\varepsilon > 0$ and $\delta \in (0,1)$, a randomized mechanism $\mathcal{M}$ provides event-level $(\varepsilon, \delta)$-differential privacy, or $(\varepsilon, \delta)$-DP for short, if for any two neighboring streams $S, S'$ and all sets of outputs $O \subseteq \text{Range}(\mathcal{M})$, $\Pr[\mathcal{M}(S) \in O] \leq \exp(\varepsilon) \cdot \Pr[\mathcal{M}(S') \in O] + \delta$.
The notion of $(\varepsilon, \delta)$-DP guarantees that the outputs of $\mathcal{M}$ on $S$ and $S'$ are almost indistinguishable, as quantified by $\varepsilon$ and $\delta$: the closer they are to $0$, the lower the distinguishability and the higher the protection level, and vice versa.

For simplicity of privacy analysis, we also consider an alternative notion of $(\varepsilon, \delta)$-DP, namely $\rho$-zero-concentrated differential privacy ($\rho$-zCDP) \cite{BunS16}.
Specifically, a mechanism $\mathcal{M}$ satisfies $\rho$-zCDP if for any two neighboring streams $S, S'$ and all $\alpha \in (1, \infty)$, $D_{\alpha}(\mathcal{M}(S) \| \mathcal{M}(S')) \leq \rho \alpha$, where $D_{\alpha}(\cdot \| \cdot)$ is the $\alpha$-R\'enyi divergence of two distributions.
The definition of $\rho$-zCDP satisfies several key characteristics of differential privacy for analysis:
(1) \emph{adaptive composition}: if $\mathcal{M}$ and $\mathcal{M}'$ satisfy $\rho$-zCDP and $\rho'$-zCDP, their composite mechanism $\mathcal{M}''(S) = (\mathcal{M}(S), \mathcal{M}'(S))$ will satisfy $(\rho + \rho')$-zCDP; (2) \emph{parallel composition}: for a $\rho$-zCDP mechanism $\mathcal{M}$ that operates on disjoint substreams $S^{(1)}, \dots, S^{(l)}$ of $S$ such that $S^{(1)} \cup \dots \cup S^{(l)} = S$, the union of the outputs of $\mathcal{M}(S^{(1)}), \dots, \mathcal{M}(S^{(l)})$ will satisfy $\rho$-zCDP; (3) \emph{post-processing immunity}: for a $\rho$-zCDP mechanism $\mathcal{M}$ and any function $g$ taking the output of $\mathcal{M}$ as input, a new mechanism $\mathcal{M}'(S) = g(\mathcal{M}(S))$ satisfies $\rho$-zCDP.
The relationship between $\rho$-zCDP and $(\varepsilon, \delta)$-DP is given below.
\begin{theorem}[zCDP $\Rightarrow$ DP]\label{thm-zCDP}
  If $\mathcal{M}$ provides $\rho$-zCDP, $\mathcal{M}$ will also provide $(\rho + 2 \sqrt{\rho \ln(1/\delta)}, \delta)$-DP for any $\delta \in (0, 1)$.
\end{theorem}

A prototypical example of a mechanism satisfying $\rho$-zCDP is the Gaussian mechanism \cite{DworkKMMN06}.
It perturbs a real-valued result by injecting Gaussian noise, the scale of which depends on the $l_2$-sensitivity of the result.
A function $g: \mathbb{N}^{m} \mapsto \mathbb{R}^d$ has $l_2$-sensitivity $\Delta_{2}$ if for all neighboring $S, S'$, $\|g(S) - g(S')\|_2 \leq \Delta_{2}$.
For a function $g: \mathbb{N}^{m} \mapsto \mathbb{R}^d$ with $l_2$-sensitivity $\Delta_{2}$, the Gaussian mechanism is defined as $\mathcal{M}_{G}(S) = g(S) + \mathcal{N}(0, \sigma^2 \cdot \mathbb{I}_d)$, where $\mathcal{N}(0, \sigma^2 \cdot \mathbb{I}_d)$ is a $d$-dimensional Gaussian random variable with mean zero and covariance matrix $\sigma^2 \cdot \mathbb{I}_d$ and $\mathbb{I}_d$ is the identity matrix.
For any $\rho > 0$, the Gaussian mechanism with $\sigma^2 = \Delta_2^2/(2\rho)$ satisfies $\rho$-zCDP \cite{BunS16}.

\paragraph{(Private) Count-Min Sketch}
The Count-Min Sketch (CMS) \cite{CormodeM05} is a common data structure for frequency estimation.
It consists of a two-dimensional array of counters, denoted as $\mathcal{C}[a, b]$, where each counter is initialized to $0$, and a set of $a$ independent hash functions $h_1, \dots, h_a$, each of which maps the items in $\mathcal{E}$ uniformly to $[b]$.
Upon receiving an item $s_i$, it increments $a$ counters by $1$, one per each row, based on the hash values of $s_i$, i.e., $\mathcal{C}[r, h_r(s_i)] \gets \mathcal{C}[r, h_r(s_i)] + 1$ for each $r \in [a]$.
After processing all $n$ items in $S$, to obtain the frequency of $e \in \mathcal{E}$, a CMS computes the hash value $h_r(e)$ for each $r \in [a]$, retrieves the corresponding counter $\mathcal{C}[r, h_r(e)]$, and takes the minimum value, i.e., $\tilde{f}_e = \min_{r \in [a]} \mathcal{C}[r, h_r(e)]$.
Theoretically, when $a = O(\log\frac{1}{\eta})$ and $b = O(\frac{1}{\zeta})$ for the parameters $\zeta, \eta \in (0, 1)$, a CMS uses $O(\frac{1}{\zeta}\log\frac{1}{\eta})$ space, takes $O(\log\frac{1}{\eta})$ time to process each update and frequency query, and guarantees that $0 \leq \tilde{f}_e - f_e \leq \zeta n$ with probability at least $1 - \eta$.

\begin{algorithm}[t]
    \small
    \caption{Private Count-Min Sketch \cite{ZhaoQRAAW22}}
    \label{alg:pcms}
    \begin{algorithmic}[1]
        \Require{Stream $S$, sketching parameters $\zeta, \eta \in (0, 1)$, private budget $\rho > 0$}
        \LComment{PCMS construction}
        \State Initialize an array $\widehat{\mathcal{C}}$ of size $a \times b$, where $a = O\big(\log\frac{1}{\eta}\big)$ and $b = O(\frac{1}{\zeta})$, with random variables drawn independently from $\mathcal{N}(0, \frac{a}{\rho})$
        \For{$i = 1$ \textbf{to} $n$}
            \For{$r = 1$ \textbf{to} $a$\label{ln-add-s}}
                \State $\widehat{\mathcal{C}}[r, h_r(s_i)] \gets \widehat{\mathcal{C}}[r, h_r(s_i)] + 1$\label{ln-add-t}
            \EndFor
        \EndFor
        \State \Return $\widehat{\mathcal{C}}$
        \LComment{Frequency query on item $e \in \mathcal{E}$ using PCMS}
        \For{$r = 1$ \textbf{to} $a$\label{ln-cms-query-s}}
            \State Compute $h_r(e)$ and obtain $\widehat{\mathcal{C}}[r, h_r(e)]$
        \EndFor
        \State \Return $\widehat{f}_e = \min_{r \in [a]} \widehat{\mathcal{C}}[r, h_r(e)]$\label{ln-cms-query-t}
    \end{algorithmic}
\end{algorithm}

A CMS can be modified to be differentially private \cite{ZhaoQRAAW22} based on the notion of $\rho$-zCDP \cite{BunS16} and the Gaussian mechanism \cite{DworkKMMN06}.
The private CMS (PCMS) follows the same structure and procedures for the construction and frequency queries as the original CMS, as presented in Algorithm~\ref{alg:pcms}.
In the construction process, the only difference is that the PCMS initializes each counter with a random variable drawn from the Gaussian distribution $\mathcal{N}(0, \sigma^2)$.
Since the $l_2$-sensitivity of the counters in a CMS is $\Delta_2 = \sqrt{2a}$, as each item leads to at most $2a$ changes in the counters, the variance of the Gaussian noise added to each counter is $\sigma^2 = a/\rho$ to ensure $\rho$-zCDP.
According to \cite{ZhaoQRAAW22}, with the same $a = O(\log\frac{1}{\eta})$ and $b = O(\frac{1}{\zeta})$, the computational and space efficiencies of the PCMS remain the same as the original CMS.
The PCMS has higher errors in frequency estimation than the original CMS due to Gaussian noise, as given in the following theorem.
\begin{theorem}\label{theorem:modified-private-count-min}[cf.~\cite[Theorem~3.1]{ZhaoQRAAW22}]
  The estimated frequency $\widehat{f}_e$ of each $e \in \mathcal{E}$ returned by a PCMS $\widehat{\mathcal{C}}$ satisfies $\lvert \widehat{f}_e - f_e \rvert \leq \zeta n + \xi$, where $\xi = \sqrt{\frac{2}{\rho} \log{\frac{2}{\eta}}} \cdot \sqrt{\log\big(\frac{4}{\zeta \eta} \log\frac{2}{\eta}\big)}$, with probability at least $1 - \eta$.
\end{theorem}

\paragraph{Problem Formulation}
We aim to maintain the frequency of any item in $\mathcal{E}$ and the set of $\gamma$-heavy hitters at any time over a size-$w$ sliding window with $(\varepsilon, \delta)$-DP.
Due to the randomized nature of sketches and DP mechanisms, providing exact results for both queries is infeasible in our setting.
Therefore, we consider building a (sublinear) data structure (i.e., \emph{sketch}) to provide approximate results for frequency and heavy-hitter queries, as defined below.
\begin{definition}[$(\xi, \eta)$-Approximate Frequency]\label{def-prob-freq}
  Given any $\xi > 0$ and $\eta \in (0, 1)$, if $\Pr[|\widehat{f}_{e, t} - f_{e, t}| \leq \xi] \geq 1 - \eta$, we call $\widehat{f}_{e, t}$ a $(\xi, \eta)$-approximation for the frequency $f_{e, t}$ of an item $e \in \mathcal{E}$ in $W_t$.
\end{definition}
\begin{definition}[$(\xi, \eta)$-Approximate $\gamma$-Heavy Hitters \cite{CormodeM05,abs-2302-11081}]\label{def-prob-hh}
  Given any $\xi > 0$ and $\eta \in (0, 1)$, a set $\widehat{\mathcal{H}}_{\gamma, t} \subseteq \mathcal{E}$ is a $(\xi, \eta)$-approximation for the set $\mathcal{H}_{\gamma, t}$ of $\gamma$-heavy hitters in $W_t$ if for each item $e \in \mathcal{E}$ with $f_{e, t} \geq \gamma w + \xi$, we have $e \in \widehat{\mathcal{H}}_{\gamma, t}$ with probability at least $1 - \eta$, and for each item $e \in \mathcal{E}$ with $f_{e, t} < \frac{\gamma w}{2} - \xi$, we have $e \notin \widehat{\mathcal{H}}_{\gamma, t}$ with probability at least $1 - \eta$.
\end{definition}

\begin{table}[t]
    \small
    \caption{List of Frequently Used Notations}
    \label{tab-notation}
    \vspace{-1em}
    \centering
    \begin{tabular}{cl}
        \toprule
        \textbf{Symbol} & \textbf{Definition}\\
        \midrule
        $S$; $n$ & Data stream; size of the stream, i.e., $n = |S|$\\
        $\mathcal{E}$; $m$ & Domain of items; domain size, i.e., $m = |\mathcal{E}|$\\
        $S_t$ & Prefix of $S$ until the arrival of $s_t$ at time $t$\\
        $W_t$; $w$ & Sliding window at time $t$; window size\\
        $f_{e, t}$ & Frequency of item $e \in \mathcal{E}$ in $W_t$\\
        $\mathcal{H}_{\gamma, t}$ & Set of $\gamma$-heavy hitters in $W_t$ for a threshold $\gamma \in (0, 1)$\\
        $\varepsilon, \delta$ & Privacy parameters in DP\\
        $\rho$ & Privacy parameters in zCDP\\
        $\zeta, \eta$ & Error and confidence parameters in CMS and PCMS\\
        $\widehat{f}_{e, t}$; $\widehat{\mathcal{H}}_{\gamma, t}$ & Approximations of $f_{e, t}$ and $\mathcal{H}_{\gamma, t}$ with DP\\
        $\xi$ & Error parameter in the approximations of $f_{e, t}$ and $\mathcal{H}_{\gamma, t}$\\
        $\mathcal{C}$; $\widehat{\mathcal{C}}$ & (Non-private) CMS; PCMS\\
        \bottomrule
    \end{tabular}
\end{table}

Before discussing technical details, we summarize the frequently used notations in Table~\ref{tab-notation}.

\section{The DPSW-Sketch Framework}
\label{sec-method}

In this section, we propose the \textsc{DPSW-Sketch} framework to privately maintain the two approximate sliding-window statistics in Definitions~\ref{def-prob-freq} and~\ref{def-prob-hh}.
We first describe how \textsc{DPSW-Sketch} is built and used for query processing in Section~\ref{subsec-alg}.
Next, we analyze its privacy guarantee, utility bound, and complexity in Section~\ref{subsec-theory}.

\begin{figure}[t]
  \centering
  \includegraphics[width=.75\linewidth]{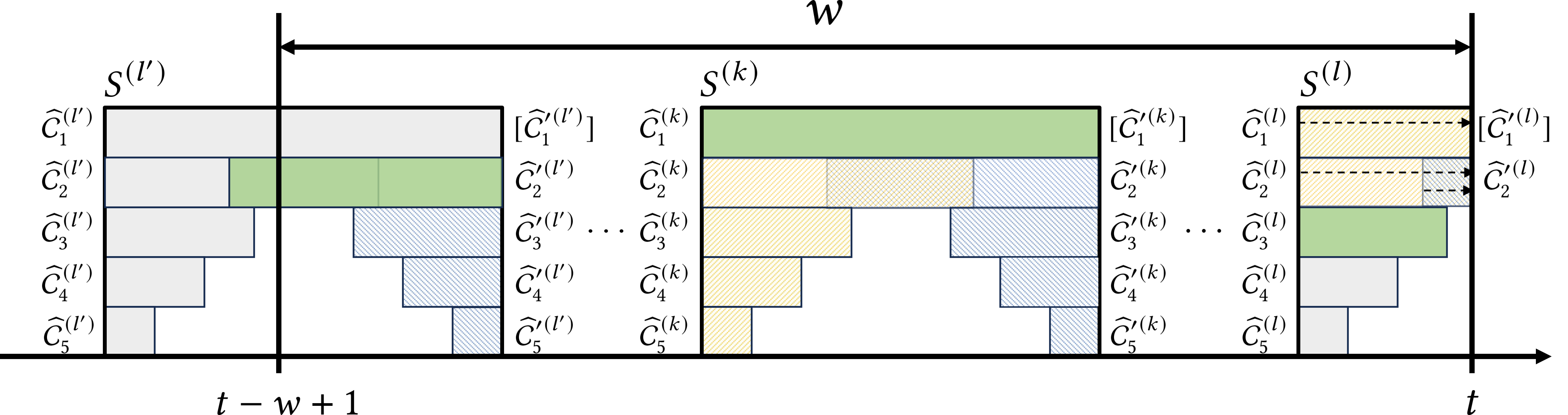}
  \vspace{-1em}
  \caption{Overview of the \textsc{DPSW-Sketch} framework.}
  \label{fig:construct}
  \Description{framework}
\end{figure}

\subsection{Algorithm Description}
\label{subsec-alg}

We illustrate the structure of the \textsc{DPSW-Sketch} framework in Figure~\ref{fig:construct}.
Generally, \textsc{DPSW-Sketch} partitions the prefix $S_t$ of the stream $S$ at any time $t$ into non-overlapping substreams $S^{(1)}, \dots,$ $S^{(l)}$ of equal length.
Then, based on the idea of \emph{smooth histogram} \cite{BravermanO07}, we establish two lists of checkpoints in each substream and build a PCMS w.r.t.~each checkpoint.
For one list (in yellow), a PCMS is built from the beginning of the substream to each checkpoint.
For the other list (in blue), a PCMS is constructed from each checkpoint to the end of the substream.
To reduce space usage, we only keep PCMSs built in substreams that overlap with $W_t$, as expired items are irrelevant to queries on $W_t$.
Finally, to process any query, we combine the results obtained from one PCMS per substream (in green), so that the concatenation of the items on which they are built is closest to $W_t$.

\begin{algorithm}[t]
    \small
    \caption{\textsc{DPSW-Sketch} Construction}
    \label{alg:construct}
    \begin{algorithmic}[1]
    \Require{Stream $S$, window size $w \in \mathbb{Z}^+$, privacy budget $\rho > 0$, framework factors $\alpha, \beta \in (0, 1)$, sketching parameters $\zeta, \eta \in (0, 1)$}
    \Ensure{\textsc{DPSW-Sketch} $\widehat{\mathcal{C}}_t$ at time $t$}
    \State Initialize $I, I' \gets \emptyset$ and $l \gets 0$
    \LComment{Pre-specify the indices of checkpoints in each substream}
    \For{$i = \lceil w^{\beta} \rceil$ \textbf{to} $1$}
        \State Append a new index $i$ to the end of $I$
        \For{$j = 1$ \textbf{to} $|I| - 2$}
            \State Find the largest $k > j$ such that $I[k] \geq (1 - \alpha) \cdot I[j]$
            \State Delete all indices between $j$ and $k$ (exclusive) from $I$ and renumber the remaining indices accordingly
        \EndFor
    \EndFor
    \For{$j = 1$ \textbf{to} $|I|$}
        \State Append an index $\lceil w^{\beta} \rceil - I[j] + 1$ as $I'[j]$
    \EndFor
    \LComment{Sketch update w.r.t.~an item $s_t \in S$}
    \For{$t = 1$ \textbf{to} $n$}
        \If{$t = 1$ \textbf{or} $|S^{(l)}| \geq \lceil w^{\beta} \rceil$}
            \State $l \gets l + 1$ and start a new substream $S^{(l)}$
            \State Build a PCMS $\widehat{\mathcal{C}}^{(l)}_1$ with $\zeta, \eta$, and $\rho_1 = \rho (2\alpha - \alpha^2)$ between $t$ and $v^{(l)}_1 = t + \lceil w^{\beta} \rceil - 1$
            \For{$j = 2$ \textbf{to} $|I|$}
                \State Build a PCMS $\widehat{\mathcal{C}}^{(l)}_j$ with $\zeta, \eta$ and $\rho_j = \rho \alpha^{j - 2}(1-\alpha)^3 / 2$ between $t$ and $v^{(l)}_j = t + I[j] - 1$
                \State Build a PCMS $\widehat{\mathcal{C}}^{\prime(l)}_j$ with $\zeta, \eta$ and $\rho_j = \rho \alpha^{j - 2}(1-\alpha)^3 / 2$ between $u^{(l)}_j = t + I'[j]$ and $t + \lceil w^{\beta} \rceil - 1$
            \EndFor
        \EndIf
        \For{$j = 1$ \textbf{to} $|I|$}
            \State \textbf{if} $v^{(l)}_j \geq t$ \textbf{then} $\widehat{\mathcal{C}}^{(l)}_j$.add$(s_t)$
            \State \textbf{if} $j > 1$ \textbf{and} $u^{(l)}_j \leq t$ \textbf{then} $\widehat{\mathcal{C}}^{\prime(l)}_j$.add$(s_t)$
        \EndFor
        \State Find the smallest $l' < l$ with $S^{(l')} \cap W_t \neq \emptyset$ and delete all PCMSs built on $S^{(l' - 1)}$ and previous substreams
        \State Use the union of PCMSs on $S^{(l')}, \dots, S^{(l)}$ as $\widehat{\mathcal{C}}_t$ at time $t$
    \EndFor
    \end{algorithmic}
\end{algorithm}

\paragraph{Sketch Construction}
The detailed procedure to construct \textsc{DPSW-Sketch} over the sliding window is presented in Algorithm~\ref{alg:construct}.
Generally, \textsc{DPSW-Sketch} uses two factors $\alpha, \beta \in (0, 1)$ to determine its structure: $\alpha$ controls the number of checkpoints in each substream and $\beta$ specifies the size of each substream as $\lceil w^{\beta} \rceil$.
The first step in the construction process is to pre-specify a list $I$ of indices of checkpoints from $\lceil w^{\beta} \rceil$ to $1$ following the maintenance procedure of \emph{smooth histogram} with the smooth factor $\beta$ \cite{BravermanO07}.
The resultant list $I$ has two properties: (1) $I$ contains as few checkpoints as possible ($|I| = O(\frac{\log{w}}{\alpha})$ according to \cite{BravermanO07}); and (2) it guarantees that for any two neighboring checkpoints $I[k], I[k - 1]$ ($k = 2, \dots, |I|$), $I[k]$ must be either at least $(1 - \alpha) \cdot I[k - 1]$ or $I[k - 1] - 1$.
Then, we reverse each index in $I$ to obtain the other set $I'$ of indices.
As such, we use each checkpoint in $I$ or $I'$ to indicate the end point of a PCMS that starts from the beginning of a substream or the start point of a PCMS that finishes at the end of a substream, respectively, both relative to the beginning of the substream.

After obtaining $I$ and $I'$, the algorithm proceeds to maintain the sketch framework when receiving a new item $s_t$ in the stream $S$ at time $t$ ($t = 1, \dots, n$).
When the first item is received or the current substream is full, we start a new substream $S^{(l)}$, initialize two sets $\widehat{\mathcal{C}}^{(l)}, \widehat{\mathcal{C}}^{\prime(l)}$ of PCMSs within $S^{(l)}$ according to $I$ and $I'$, and allocate the private budget $\rho$ among the PCMSs.
Since the first indices in $I$ and $I'$ both correspond to the entire substream, we build only one PCMS for them.
Additionally, we allocate higher privacy budgets to PCMSs maintained on more items to reduce total estimation errors.
Subsequently, the framework is updated by adding the current item $s_t$ to all PCMSs in $\widehat{\mathcal{C}}^{(l)}$ and $\widehat{\mathcal{C}}^{\prime(l)}$ whose corresponding ranges contain $s_t$, each following Lines~\ref{ln-add-s}--\ref{ln-add-t} of Algorithm~\ref{alg:pcms}.
Finally, we check whether there is any obsolete substream, where all items come before the beginning of $W_t$ and are irrelevant to the statistics on $W_t$ and in the future, and remove all PCMSs within them from the framework.
After that, the union of all PCMSs in substreams that overlap with $W_t$ serves as the sketch $\widehat{\mathcal{C}}_t$ at time $t$.

\paragraph{Query Processing}
The methods to compute the results for frequency estimation and heavy-hitter identification on $W_t$ using \textsc{DPSW-Sketch} are presented in Algorithm~\ref{alg:query}.
To obtain the frequency of a query item $e$ in $W_t$, the algorithm finds and queries one PCMS within each substream that overlaps with $W_t$.
For any intermediate substream that is fully covered by $W_t$, the PCMS w.r.t.~the whole substream is selected.
For the first substream that might contain expired items, it chooses the PCMS that starts before, but is closest to, the beginning of $W_t$.
Similarly, for the current substream, it identifies the PCMS that is closest to $t$ and has already been completed.
Finally, the estimated frequency $\widehat{f}_{e,t}$ of item $e$ on $W_t$ is the sum of frequencies queried from all selected PCMSs according to Lines~\ref{ln-cms-query-s}--\ref{ln-cms-query-t} of Algorithm~\ref{alg:pcms}.
The method can also be extended to find an (approximate) set of $\gamma$-heavy hitters for a threshold $\gamma \in (0, 1)$.
It first queries the frequency $\widehat{f}_{e, t}$ of each item $e \in \mathcal{E}$ according to Lines~\ref{ln-query-s}-\ref{ln-query-t}.
Then, each item $e \in \mathcal{E}$ is checked one by one: If $\widehat{f}_{e, t}$ is greater than or equal to $(\gamma - \zeta) w$, $e$ will be added to $\widehat{\mathcal{H}}_{\gamma, t}$.
After processing all items, $\widehat{\mathcal{H}}_{\gamma, t}$ is returned for the heavy-hitter query.

\begin{algorithm}[t]
    \small
    \caption{\textsc{DPSW-Sketch} Query Processing}
    \label{alg:query}
    \begin{algorithmic}[1]
        \Require{\textsc{DPSW-Sketch} $\widehat{\mathcal{C}}_t$ at time $t$, window size $w \in \mathbb{Z}^+$, item $e \in \mathcal{E}$ (for frequency query) or threshold $\gamma \in (0, 1)$ (for heavy-hitter query)}
        \Ensure{(Approximate) frequency $\widehat{f}_{e,t}$ or set $\widehat{\mathcal{H}}_{\gamma, t}$ of $\gamma$-heavy hitters}
        \LComment{Frequency query on item $e \in \mathcal{E}$ using \textnormal{\textsc{DPSW-Sketch}}}
        \State $x \gets \argmax_{1 \leq j \leq |I|} u^{(l')}_{j} \leq \max(t - w + 1, 1)$ \label{ln-query-s}
        \State $y \gets \argmax_{1 \leq j \leq |I|} v^{(l)}_{j} \leq t$
        \State $\widehat{f}_{e}^{(l')} \gets \widehat{\mathcal{C}}^{\prime(l)}_x$.query$(e)$ and $\widehat{f}_{e}^{(l)} \gets \widehat{\mathcal{C}}^{(l)}_y$.query$(e)$
        \For{$k = l' + 1$ \textbf{to} $l - 1$}
            \State $\widehat{f}_{e}^{(k)} \gets \widehat{\mathcal{C}}^{(k)}_1$.query$(e)$
        \EndFor
        \State \Return $\widehat{f}_{e,t} \gets \sum_{k = l'}^{l} \widehat{f}_{e}^{(k)}$ \label{ln-query-t}
        \LComment{Heavy-hitter query using \textnormal{\textsc{DPSW-Sketch}}}
        \State Initialize $\widehat{\mathcal{H}}_{\gamma, t} \gets \emptyset$
        \ForAll{$e \in \mathcal{E}$}
            \State Compute $\widehat{f}_{e,t}$ as Lines~\ref{ln-query-s}--\ref{ln-query-t}
            \State \textbf{if} $\widehat{f}_{e,t} \geq (\gamma - \zeta) w$ \textbf{then} $\widehat{\mathcal{H}}_{\gamma, t} \gets \widehat{\mathcal{H}}_{\gamma, t} \cup \{e\}$
        \EndFor
        \State \Return $\widehat{\mathcal{H}}_{\gamma, t}$
    \end{algorithmic}
\end{algorithm}

\subsection{Theoretical Analysis}
\label{subsec-theory}

We first provide a formal privacy guarantee for \textsc{DPSW-Sketch}.
\begin{lemma}
\label{lm:privacy}
    \textnormal{\textsc{DPSW-Sketch}} $\widehat{\mathcal{C}}_t$ at any time $t$ constructed by Algorithm~\ref{alg:construct} satisfies $\rho$-zCDP.
\end{lemma}
\begin{proof}
    According to \cite[Theorem~3.1]{ZhaoQRAAW22}, each PCMS in $\widehat{\mathcal{C}}_t$ satisfies zCDP w.r.t.~the privacy parameter with which it is initialized.
    That is, for any $k \geq 1$, $\widehat{\mathcal{C}}^{(k)}_1$ provides $\rho_1$-zCDP, where $\rho_1 = \rho (2\alpha - \alpha^2)$; $\widehat{\mathcal{C}}^{(k)}_j$ and $\widehat{\mathcal{C}}^{\prime(k)}_j$ provide $\rho_j$-zCDP, where $\rho_j = \rho \alpha^{j - 2}(1-\alpha)^3 / 2$, for each $j = 2, \ldots, |L|$.
    Since all of these PCMSs are built in the same substream $S^{(k)}$, their total privacy cost $\rho_{all}$ is calculated by applying the adaptive composition of zCDP as:
    \begin{equation*}
        \rho_{all} = \rho (2\alpha - \alpha^2) + 2 \sum_{j = 2}^{|L|} \frac{\rho}{2} \alpha^{j-2}(1-\alpha)^3
        = \rho - \rho(1 - \alpha)^2 + \rho(1 - \alpha)^{2}(1 - \alpha^{|L| - 1}) < \rho.
    \end{equation*}
    Consequently, the output in each substream consisting of the union of all PCMSs satisfies $\rho$-zCDP.
    Finally, since all substreams are mutually disjoint, \textsc{DPSW-Sketch} $\widehat{\mathcal{C}}_t$ at any time $t$ built by Algorithm~\ref{alg:construct} satisfies $\rho$-zCDP due to the parallel composition of zCDP.
\end{proof}
According to \cite{ZhaoQRAAW22}, a PCMS can only guarantee zCDP for queries. That is, it assumes that an adversary has no access to the internal sketch structure and is allowed to query a PCMS \emph{after} the PCMS has been built on all items in the dataset.
As such, any query can be regarded as a \emph{post-processing} step without additional privacy losses.
\textsc{DPSW-Sketch} follows the same setting as \cite{ZhaoQRAAW22}: By dividing the stream into disjoint substreams and using smooth histograms to pre-specify the range in which each PCMS is built, we ensure that all PCMSs are queried only after they have been built without updates.
Therefore, \textsc{DPSW-Sketch} can answer an unlimited number of queries at all times with DP.
We consider how to convert $\rho$-zCDP to $(\varepsilon, \delta)$-DP according to Theorem~\ref{thm-zCDP}.
Specifically, to guarantee $(\varepsilon, \delta)$-DP for any $\varepsilon > 0$ and $\delta \in (0, 1)$, the value of $\rho$ is set as:
\begin{equation}\label{eq-rho}
    \rho = \varepsilon +2 \ln(1/\delta) - 2 \sqrt{\varepsilon \ln(1/\delta) + \ln^2(1/\delta)}.
\end{equation}

Next, we analyze the error of the \textsc{DPSW-Sketch} framework for frequency estimation.
\begin{lemma}
\label{lm-freq}
    The frequency $\widehat{f}_{e, t}$ by Algorithm~\ref{alg:query} satisfies $\Pr\big[|\widehat{f}_{e, t} - f_{e, t}| \leq \zeta w + \xi\big] \geq 1 - O(\sqrt{w} \eta)$, where $\xi = O\big(\sqrt{w}(\zeta + \frac{1}{\sqrt{\rho}} \log\frac{1}{\zeta \eta})\big)$.
\end{lemma}
\begin{proof}
    We compute the error bound of $\widehat{f}_{e, t}$ for estimating $f_{e, t}$ by summing the upper bounds of its errors in all substreams.
    For each substream $S^{(k)}$ ($l' < k < l$) that is fully covered by $W_t$, we always obtain $\widehat{f}^{(k)}_{e}$ from $\widehat{\mathcal{C}}^{(k)}_{1}$.
    According to Theorem~\ref{theorem:modified-private-count-min}, we have
    \begin{equation}\label{eq-err1}
        \Pr\big[ |\widehat{f}^{(k)}_{e} - f^{(k)}_{e}| \leq \zeta \cdot \lceil w^{\beta} \rceil + \xi^{(k)} \big] \geq 1 - \eta,
    \end{equation}
    where $\xi^{(k)} = \sqrt{\frac{2 \log(2/\eta)}{\rho(2\alpha - \alpha^2)}} \cdot \sqrt{\log\big(\frac{4 \log(2/\eta)}{\zeta \eta}\big)}$.
    For $S^{(l')}$ and $S^{(l)}$ that partially overlap with $W_t$, the errors come from the estimates of PCMSs and the misalignments between $W_t$ and the ranges of the PCMSs used for query processing.
    Specifically, we have
    \begin{gather}
        \Pr\big[ |\widehat{f}^{(l')}_{e} - f^{(l')}_{e}| \leq (\alpha + \zeta) \cdot I[x] + \xi^{(l')} \big] \geq 1 - \eta; \label{eq-err2}\\
        \Pr\big[ |\widehat{f}^{(l)}_{e} - f^{(l)}_{e}| \leq (\alpha + \zeta) \cdot I[y] + \xi^{(l)} \big] \geq 1 - \eta, \label{eq-err3}
    \end{gather}
    where $\xi^{(l')} = \sqrt{\frac{2 \log(2/\eta)}{\rho_x}} \cdot \sqrt{\log\big(\frac{4 \log(2/\eta)}{\zeta \eta}\big)}$
    and $\xi^{(l)} = \sqrt{\frac{2 \log(2/\eta)}{\rho_y}} \cdot \sqrt{\log\big(\frac{4 \log(2/\eta)}{\zeta \eta}\big)}$.
    Since there are at most $l - l' + 1 \leq \lceil w^{1-\beta} \rceil + 1$ active substreams, by combining Eqs.~\ref{eq-err1}--\ref{eq-err3} and applying the union bound and the triangle inequality, we obtain with probability at least $1 - (\lceil w^{1-\beta} \rceil + 1) \eta$,
    \begin{equation}\label{eq-err4}
        |\widehat{f}_{e, t} - f_{e, t}| \leq \sum_{k = l'}^{l} |\widehat{f}^{(k)}_{e} - f^{(k)}_{e}| \leq \sum_{k = l' + 1}^{l - 1} \Big(\zeta \lceil w^{\beta} \rceil + \xi^{(k)} \Big) + (\alpha + \zeta) (I[x] + I[y]) + \xi^{(l')} + \xi^{(l)}.
    \end{equation}
    Taking $\beta = \frac{1}{2}$ and $\alpha = O(1)$, Eq.~\ref{eq-err4} can be simplified as
    \begin{align*}
        |\widehat{f}_{e, t} - f_{e, t}| \leq \zeta w + O\Big(\sqrt{w} \big(\zeta + \frac{1}{\sqrt{\rho}} \log\frac{1}{\zeta \eta}\big)\Big),
    \end{align*}
    because $\xi^{(k)} = O\big(\frac{1}{\sqrt{\rho}} \log\frac{1}{\zeta \eta} \big)$ for each $k \in [l', \dots, l]$ and $I[x] = I[y] = O(\sqrt{w})$, and the proof is concluded.
\end{proof}

Then, we extend the error bound to heavy-hitter identification.
\begin{lemma}
\label{lm-heavy-hitter}
    The set $\widehat{\mathcal{H}}_{\gamma, t}$ returned by Algorithm~\ref{alg:query} satisfies (1) for any item $e \in \mathcal{E}$ with $f_{e, t} \geq \gamma w + O\big(\sqrt{w}(\zeta + \frac{1}{\sqrt{\rho}} \log\frac{1}{\zeta \eta})\big)$, $e \in \widehat{\mathcal{H}}_{\gamma, t}$ with probability at least $1 - O(m \sqrt{w} \eta)$; (2) for any item $e \in \mathcal{E}$ with $f_{e, t} < (\gamma - 2 \zeta) w - O\big(\sqrt{w}(\zeta + \frac{1}{\sqrt{\rho}} \log\frac{1}{\zeta \eta})\big)$, $e \notin \widehat{\mathcal{H}}_{\gamma, t}$ with probability at least $1 - O(m \sqrt{w} \eta)$.
\end{lemma}
\begin{proof}
    According to Lemma~\ref{lm-freq}, for any item $e \in \mathcal{E}$ with $f_{e, t} \geq \gamma w + \xi$, where $\xi = O\big(\sqrt{w}(\zeta + \frac{1}{\sqrt{\rho}} \log\frac{1}{\zeta \eta})\big)$, we have $\widehat{f}_{e, t} \geq f_{e, t} - \zeta w - \xi \geq (\gamma - \zeta) w$ and $e$ is included in $\widehat{\mathcal{H}}_{\gamma, t}$ with probability $1 - O(\sqrt{w} \eta)$.
    Similarly, for any item $e \in \mathcal{E}$ with $f_{e, t} < (\gamma - 2 \zeta) w - \xi$, we have $\widehat{f}_{e, t} \leq f_{e, t} + \zeta w + \xi < (\gamma - \zeta) w$ and $e$ is not included in $\widehat{\mathcal{H}}_{\gamma, t}$ with probability $1 - O(\sqrt{w} \eta)$.
    By applying the union bound w.r.t.~all the $m$ items in $\mathcal{E}$, it guarantees that the above results hold for all items with probability at least $1 - O(m \sqrt{w} \eta)$.
\end{proof}

Finally, the main theoretical results of the \textsc{DPSW-Sketch} framework are summarized as follows.
\begin{theorem}
\label{thm-main}
    For any window size $w$, framework factors $\beta = \frac{1}{2}$ and $\alpha = O(1)$, sketching parameters $\zeta, \eta \in (0, 1)$, privacy parameters $\varepsilon > 0$ and $\delta \in (0, 1)$, the following results hold:
    \begin{enumerate}[label*=\arabic*.]
        \item \textnormal{\textsc{DPSW-Sketch}} is $(\varepsilon, \delta)$-differentially private.
        \item It provides an $\Big(O\big(\zeta w + \frac{\sqrt{w\log(1/\delta)}}{\varepsilon} \log\frac{w m}{\zeta \eta}\big), \eta\Big)$-approximate frequency $\widehat{f}_{e, t}$ for each item $e \in \mathcal{E}$.
        \item It gives an $\Big(O\big(\zeta\sqrt{w} + \frac{\sqrt{w\log(1/\delta)}}{\varepsilon} \log\frac{w m}{\zeta \eta}\big), \eta \Big)$-approximate set $\widehat{\mathcal{H}}_{\gamma, t}$ of $\gamma$-heavy hitters for any threshold $\gamma \geq 4\zeta$.
        \item It uses $O(\frac{\sqrt{w}\log{w}}{\zeta} \log\frac{w m}{\eta})$ space and $O(\log{w} \log\frac{w m}{\eta})$ time per update and processes a frequency query and a heavy-hitter query in $O(\sqrt{w} \log\frac{w m}{\eta})$ and $O(m \sqrt{w}$ $\log\frac{w m}{\eta})$ time.
    \end{enumerate}
\end{theorem}
\begin{proof}
    First of all, by setting the value of $\rho$ according to Eq.~\ref{eq-rho} for any given $\varepsilon > 0$ and $\delta \in (0, 1)$, \textsc{DPSW-Sketch} satisfies $(\varepsilon, \delta)$-DP.
    Eq.~\ref{eq-rho} implies $\frac{1}{\rho} = \frac{\varepsilon +2 \ln(1/\delta) + 2 \sqrt{\varepsilon \ln(1/\delta) + \ln^2(1/\delta)}}{\varepsilon^2} = O\big(\frac{\log(1 / \delta)}{\varepsilon^2}\big)$.
    By replacing $\frac{1}{\rho}$ with $O\big(\frac{\log(1 / \delta)}{\epsilon^2}\big)$ and scaling $\eta$ to $O(\frac{\eta}{m \sqrt{w}})$ for each PCMS in Lemmas~\ref{lm-freq} and~\ref{lm-heavy-hitter}, we obtain the approximation bounds for frequency estimation and heavy-hitter identification.

    In terms of space complexity, we can see that there exist $O(\sqrt{w})$ active substreams that overlap with $W_t$ for any $t \in \mathbb{Z}^{+}$, the number of PCMSs maintained in each active substream is $O(\log w)$, and the size of each PCMS is $O(\frac{1}{\zeta} \log\frac{w m}{\eta})$.
    In total, the space complexity of \textsc{DPSW-Sketch} is $O(\frac{\sqrt{w}\log{w}}{\zeta} \log\frac{w m}{\eta})$.
    In terms of time complexity, we can see that Algorithm~\ref{alg:construct} uses $O(\sqrt{w})$ time to decide the two lists of checkpoints in a smooth histogram, initializes $O(\log w)$ PCMSs at the beginning of each substream, inserts an item into at most $O(\log w)$ PCMSs, and requires $O(\log \frac{w m}{\eta})$ hash computations and counter updates for each insertion. Since the checkpoints are computed only once before stream processing, the (amortized) time complexity per update is $O(\log{w} \log\frac{w m}{\eta})$.
    For a frequency query, Algorithm~\ref{alg:query} selects $O(\sqrt{w})$ PCMSs from \textsc{DPSW-Sketch} and takes $O(\log \frac{w m}{\eta})$ time to query each PCMS. Thus, the time complexity of each frequency query is $O(\sqrt{w} \log\frac{w m}{\eta})$. A heavy-hitter query requires to perform $m$ frequency queries and thus has a time complexity of $O(m \sqrt{w} \log\frac{w m}{\eta})$.
\end{proof}

\paragraph{Comparison with Existing Private Sketches in \cite{ChanLSX12, Upadhyay19, abs-2302-11081}}
We show how the theoretical results of \textsc{DPSW-Sketch} improve over those of the existing private sliding-window sketches \cite{ChanLSX12, Upadhyay19, abs-2302-11081}.
First, the sketches in \cite{ChanLSX12, Upadhyay19, abs-2302-11081} are all specific to the identification of heavy hitters and the estimation of their frequencies with bounded errors, while \textsc{DPSW-Sketch} can approximate the frequencies of all items.
Second, \textsc{DPSW-Sketch} either provides better approximations for heavy-hitter identification or has a lower space complexity than the sketches in \cite{ChanLSX12, Upadhyay19, abs-2302-11081}.
Compared to \textsc{PCC-MG} in \cite{ChanLSX12}, which requires $O(w)$ space for an additive error of $O(\sqrt{w})$, and \textsc{U-Sketch} in \cite{Upadhyay19}, which has an additive error of $\Tilde{O}(w^{3/4})$ in $\Tilde{O}(\sqrt{w})$ space\footnote{We use $\Tilde{O}(\cdot)$ to omit all the factors poly-logarithmic w.r.t.~$w$ and independent of $w$.}, the additive error and space complexity of \textsc{DPSW-Sketch} are both $\Tilde{O}(\sqrt{w})$.
\textsc{BLMZ-Sketch} \cite{abs-2302-11081} achieves the same $\Tilde{O}(\sqrt{w})$ additive error as \textsc{DPSW-Sketch}.
However, its update time and space usage, both of which are $O(\frac{\sqrt{w}\log^4{w}}{\varepsilon \gamma})$, are much higher than those of \textsc{DPSW-Sketch}.
Finally, we note that all private sketches have higher errors and use more space than non-private sketches in the sliding window model, e.g., \cite{GouHZWLYW020, 0001JDZCH0U023}, due to the noise added to satisfy DP.

\section{Experiments}
\label{sec-experiments}

In this section, we conducted extensive experiments on five real-world and synthetic datasets to evaluate the performance of \textsc{DPSW-Sketch} compared to the state-of-the-art baselines.

\begin{table}[t]
    \small
    \caption{Dataset Information}
    \label{tab-dataset}
    \vspace{-1em}
    \centering
    \begin{tabular}{cccl}
        \toprule
        \textbf{Dataset} & $n$ & $m$ & \textbf{Description}\\
        \midrule
        AOL\tablefootnote{\url{https://www.cim.mcgill.ca/~dudek/206/Logs/AOL-user-ct-collection/}} & 10.7M & 38,411 & Web search query logs\\
        MovieLens\tablefootnote{\url{https://grouplens.org/datasets/movielens/latest/}} & 25M & 62,423 & Movie ratings\\
        WorldCup\tablefootnote{\url{https://ita.ee.lbl.gov/html/contrib/WorldCup.html}} & 1.3B & 82,592 & Web server logs\\
        Gaussian & 10M & 25,600 & Synthetic Gaussian-distributed items\\
        Zipf & 10M & 25,600 & Synthetic Zipf-distributed items\\
        \bottomrule
    \end{tabular}
\end{table}

\paragraph{Datasets}
We employed three real-world and two synthetic datasets in the experiments.
Their statistics are reported in Table~\ref{tab-dataset}, where $n$ is the dataset size and $m$ is the domain size.
For synthetic datasets, the skewness parameter for the Zipf distribution is $1$; the mean and standard deviation of the Gaussian distribution are $50$ and $25$, respectively.
More detailed information on these datasets is provided in Appendix~\ref{app-data} due to space limitations.

\paragraph{Baselines}
We adopted \textsc{U-Sketch} \cite{Upadhyay19}, \textsc{PCC-MG} \cite{ChanLSX12}, and \textsc{BLMZ-Sketch} \cite{abs-2302-11081} as baselines for frequency estimation and heavy-hitter identification with DP in the sliding window model.
We also used a non-private sliding-window sketch, \textsc{Microscope-Sketch} \cite{0001JDZCH0U023}, as a baseline to measure ``the price of privacy.''
More details about these baselines are also deferred to Appendix~\ref{app-impl}.

\paragraph{Implementation}
All the algorithms we compared were implemented in C++ and compiled with the ``-O3'' flag.
We ran each experiment with a single thread on a server with an Intel Xeon Gold 6428 processor @2.5GHz and 32GB RAM.
For each baseline, we either used the implementation published by the original authors or followed the idea of the original paper for implementation and fine-tuned the parameters.
We set the window size $w$ to $10^6$ by default.
The default parameters in \textsc{DPSW-Sketch} are as follows: (1) number of checkpoints $|I| = 3$; (2) length of each substream $0.1 w$; and (3) height $a \in [2, \dots, 5]$ and width $b \in [500, \dots, 5000]$ for each PCMS.
The parameter sensitivity of \textsc{DPSW-Sketch} is further analyzed in Appendix~\ref{app-param}.

\paragraph{Metrics}
The query workload on each dataset was generated by randomly sampling $1\%$ timestamps from time $t = w$ to $n$ and performing a set of $100$ frequency queries and a heavy-hitter query at each sampled timestamp.
We divided the items into two groups according to their ground-truth frequencies as \emph{high}- and \emph{low}-frequency items (top-$50$ most frequent vs.~all remaining items) and each set $\mathcal{Q}$ of queries has an equal number of items from each group.
The following metrics are used for performance evaluation:
\begin{itemize}
    \item \textbf{Mean Absolute Error (MAE)}, $\frac{1}{|\mathcal{Q}|} \sum_{e \in \mathcal{Q}} |\widehat{f}_{e, t} - f_{e, t}|$, is used to measure the performance for frequency queries.
    \item \textbf{Mean Relative Error (MRE)}, $\frac{1}{|\mathcal{Q}|} \sum_{e \in \mathcal{Q}} \frac{|\widehat{f}_{e, t} - f_{e, t}|}{f_{e,t}}$, is also used to measure the performance for frequency queries.
    \item \textbf{F1-score}, which is the harmonic mean of the precision score, i.e., the ratio of $|\widehat{\mathcal{H}}_{e, t} \cap \mathcal{H}_{e, t}|$ to $|\widehat{\mathcal{H}}_{e, t}|$, and the recall score, i.e., the ratio of $|\widehat{\mathcal{H}}_{e, t} \cap \mathcal{H}_{e, t}|$ to $|\mathcal{H}_{e, t}|$, is used to measure the performance for heavy-hitter queries.
    \item \textbf{Throughput}, i.e., the average number of items to insert per second (in Kops), and \textbf{Sketch Size} (in KB) are used to measure the time and space efficiency of each method.
\end{itemize}

\begin{figure}[t]
    \centering
    \includegraphics[width=.96\linewidth]{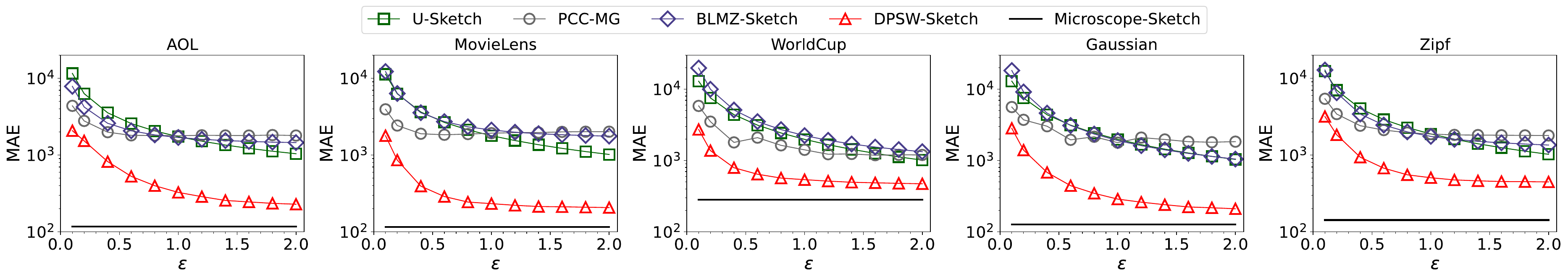}
    \includegraphics[width=.96\linewidth]{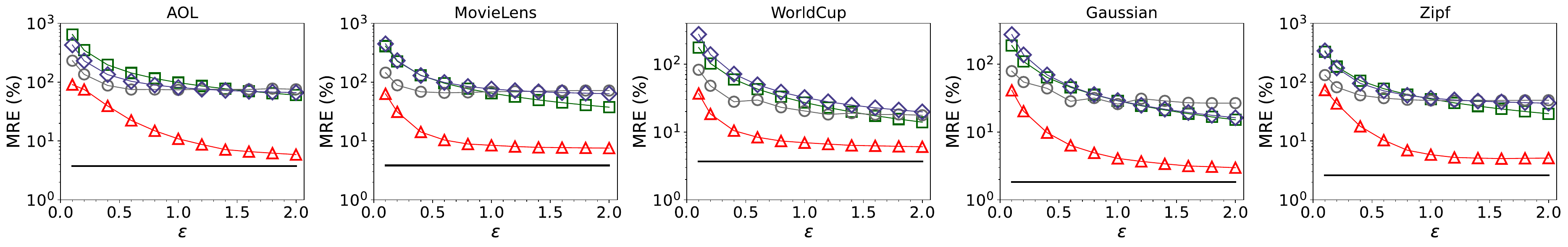}
    \vspace{-1em}
    \caption{Performance for frequency queries on high-frequency items by varying privacy parameter $\varepsilon \in \{0.1, 0.2, 0.4, \dots, 2.0\}$.}
    \label{fig-freq-eps-high}
    \Description{experiments}
\end{figure}
\begin{figure}[t]
    \centering
    \includegraphics[width=.96\linewidth]{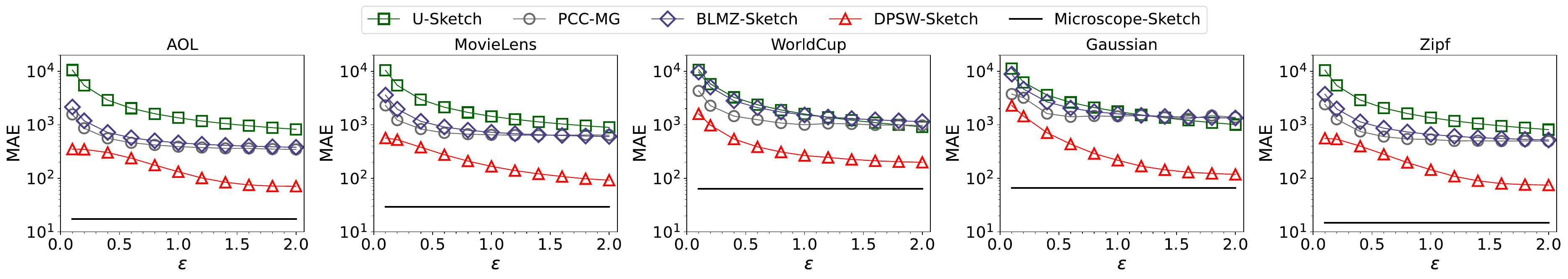}
    \includegraphics[width=.96\linewidth]{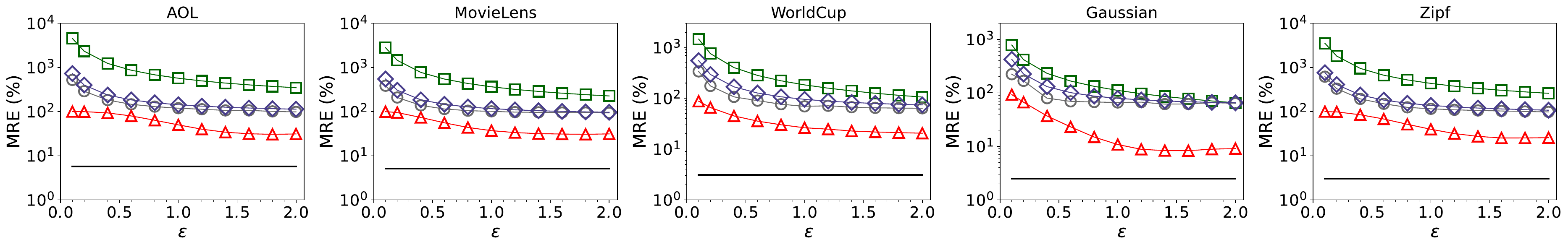}
    \vspace{-1em}
    \caption{Performance for frequency queries on low-frequency items by varying privacy parameter $\varepsilon \in \{0.1, 0.2, 0.4, \dots, 2.0\}$.}
    \label{fig-freq-eps-low}
    \Description{experiments}
\end{figure}

\subsection{Frequency Estimation}
\label{sec-exp-freq}

Figures~\ref{fig-freq-eps-high} and~\ref{fig-freq-eps-low} illustrate the performance of different sketches for frequency queries on high- and low-frequency items by varying the privacy parameter $\varepsilon$ from $0.1$ to $2$.
For sketches satisfying $(\varepsilon, \delta)$-DP and $\rho$-zCDP, we fix $\delta$ to $1/n^{1.5}$ and compute the value of $\rho$ according to Eq.~\ref{eq-rho}.
Here, we restrict the selection of low-frequency items to those with frequencies of at least 100.
This is because \textsc{PCC-MG} and \textsc{BLMZ-Sketch}, which use Misra-Gries sketches \cite{MisraG82} as building blocks, cannot answer queries on low-frequency items.
For the above query workload, \textsc{PCC-MG} and \textsc{BLMZ-Sketch} still do not report the results for many frequency queries.
In such cases, we treat their results as $0$ in the calculation of MAE and MRE.

Our results show that the MAEs and MREs of all sketches, except (non-private) \textsc{Microscope-Sketch}, generally decrease with increasing $\varepsilon$ in all datasets.
For both high- and low-frequency items, \textsc{DPSW-Sketch} achieves significantly lower MAE and MRE than all other private sketches across different values of $\varepsilon$.
Especially, \textsc{DPSW-Sketch} is the only method among them whose MREs are always within $10\%$ for high-frequency items and $100\%$ for low-frequency items when $\varepsilon \geq 1$.
The results confirm that \textsc{DPSW-Sketch} strikes a better balance between frequency estimation accuracy and privacy than state-of-the-art private sketches in the sliding-window model.

Furthermore, \textsc{U-Sketch}, \textsc{PCC-MG}, and \textsc{BLMZ-Sketch} generally exhibit performance similar to each other on high-frequency items, and \textsc{PCC-MG} is slightly better when $\varepsilon$ is smaller, while \textsc{U-Sketch} becomes better for a larger value of $\varepsilon$.
For low-frequency items, \textsc{U-Sketch} sometimes performs much worse than \textsc{PCC-MG} and \textsc{BLMZ-Sketch} because its estimation errors are even higher than simply treating all missing frequencies as $0$.
In all datasets, the three private baselines always have at least $4 \times$ higher MAE and $2 \times$ higher MRE than those of \textsc{DPSW-Sketch}.
Finally, the performance of all private sketches for frequency estimation is not comparable to that of the state-of-the-art non-private sketch, \textsc{Microscope-Sketch}, especially on low-frequency items, because (1) they introduce substantial noise in the counters to satisfy DP and (2) their sketch structures are modified for ease of privacy accounting, which further leads to estimation errors due to the misalignment between the sliding window and the sketch used for query processing.

\begin{figure}[t]
    \centering
    \includegraphics[width=.96\linewidth]{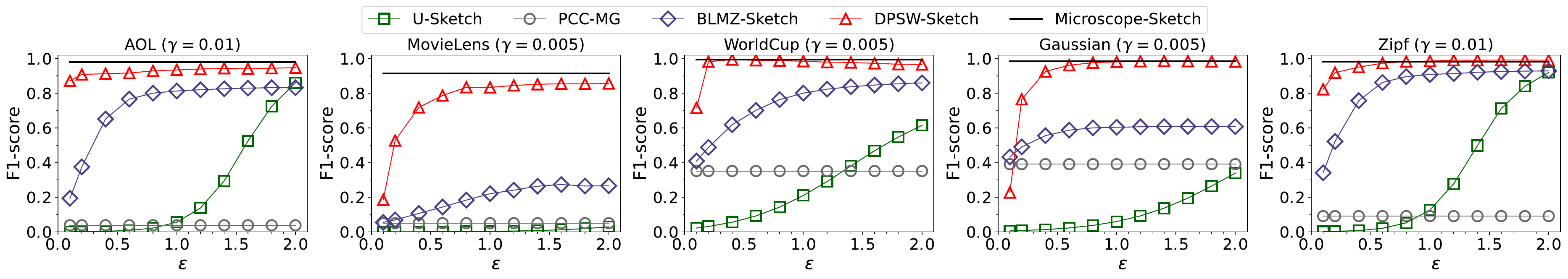}
    \includegraphics[width=.96\linewidth]{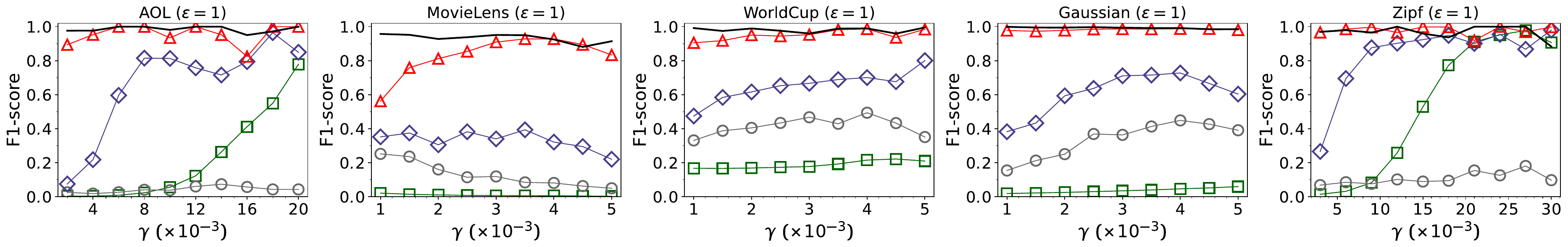}
    \vspace{-1em}
    \caption{Performance for heavy-hitter queries by varying privacy parameter $\varepsilon$ and threshold $\gamma$.}
    \label{fig-hh}
    \Description{experiments}
\end{figure}

\subsection{Heavy-Hitter Identification}
\label{sec-exp-hh}

Figure~\ref{fig-hh} shows the performance of different sketches for heavy-hitter identification by varying the privacy parameter $\varepsilon$ when the threshold $\gamma = 0.005$ or $0.01$ and the threshold $\gamma$ when $\varepsilon = 1$.
We observe that \textsc{DPSW-Sketch} achieves higher F1-scores than other DP sketches in almost all cases.
Its performance is mostly close to that of (non-private) \textsc{Microscope-Sketch} when $\varepsilon \geq 1$.
These results verify its good trade-off between accuracy for heavy-hitter identification and privacy.
The performance of \textsc{BLMZ-Sketch} and \textsc{U-Sketch} improves with increasing $\varepsilon$.
However, they are inferior to \textsc{DPSW-Sketch} due to much larger noise.
The F1-score of \textsc{PCC-MG} remains constant for all values of $\varepsilon$.
This is because \textsc{PCC-MG} is built directly on MG sketches, which only guarantee to return heavy hitters but cannot filter out non-heavy hitters.
As such, it has achieved a recall score of $1$ when $\varepsilon = 0.1$ (i.e., no false negatives) while inevitably including some false positives in all its results, regardless of the value of $\varepsilon$.
In contrast, \textsc{BLMZ-Sketch} only uses MG sketches for pre-screening and builds additional counters to decide whether a candidate item is a heavy hitter, thus achieving better performance for heavy-hitter queries.

\begin{figure}[t]
    \centering
    \includegraphics[width=.96\linewidth]{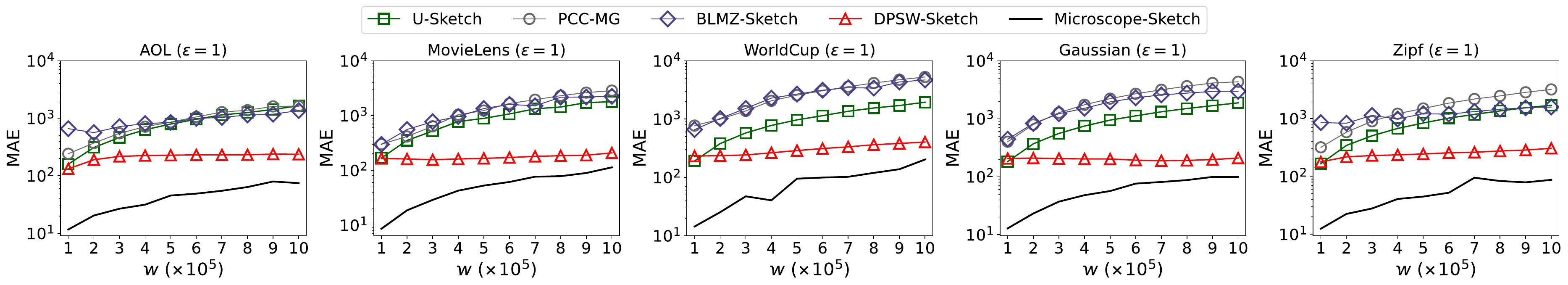}
    \includegraphics[width=.96\linewidth]{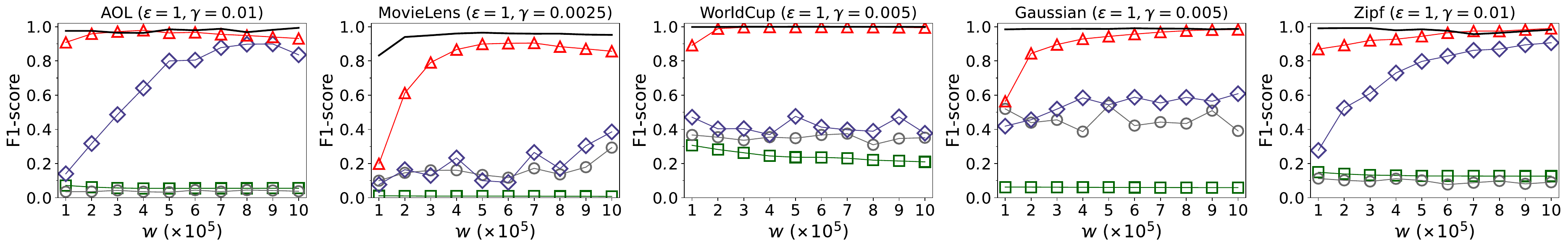}
    \vspace{-1em}
    \caption{Performance of different sketches for frequency and heavy-hitter queries by varying window size $w$.}
    \label{fig-window}
    \Description{experiments}
\end{figure}

\subsection{Performance vs.~Window Size}
\label{sec-exp-ws}

Figure~\ref{fig-window} presents the performance of different sketches for frequency and heavy-hitter queries by varying the window size $w$ from $10^5$ to $10^6$.
We observe that \textsc{DPSW-Sketch} has better query accuracies than other private sketches in almost all cases across different window sizes.
The MAEs of \textsc{DPSW-Sketch} for frequency estimation increase with the window size $w$.
This can be attributed to a greater number of distinct elements within a larger window, which leads to more hash collisions.
Note that the MAEs grow much lower than $w$ and thus the MREs of \textsc{DPSW-Sketch} drop with increasing $w$.
Furthermore, the F1-scores of \textsc{DPSW-Sketch} for heavy-hitter queries also increase with $w$.
Therefore, the performance of \textsc{DPSW-Sketch} generally degrades in smaller windows.
This is because the variance of Gaussian noise added to PCMSs is nearly independent of $w$.
As such, queries in smaller windows are more severely disturbed by noise than those in larger windows.

\begin{figure}[t]
    \centering
    \includegraphics[width=.96\linewidth]{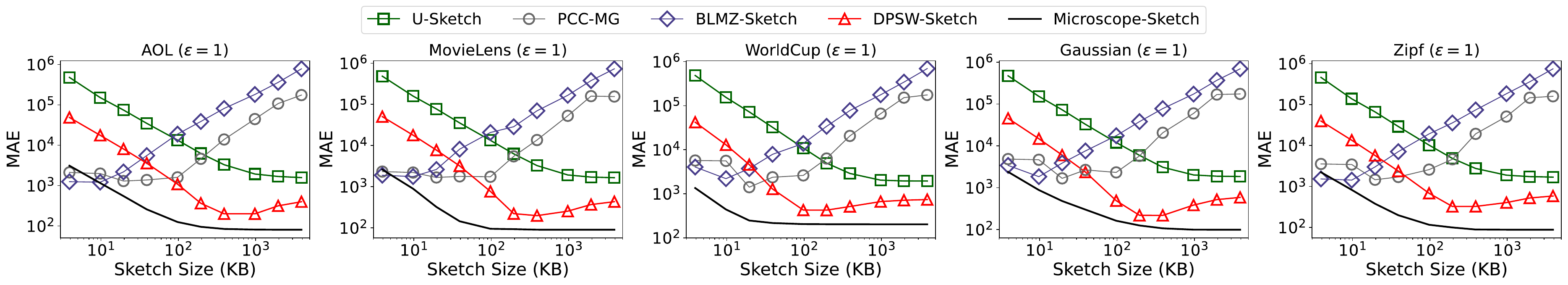}
    \includegraphics[width=.96\linewidth]{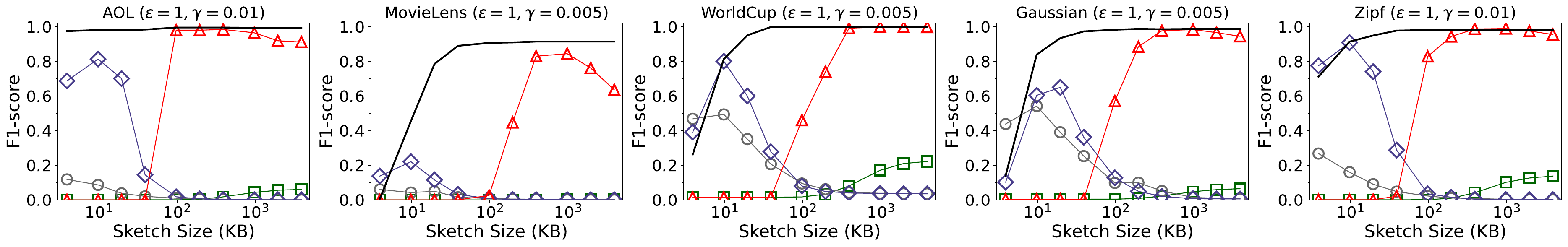}
    \vspace{-1em}
    \caption{Performance of different sketches for frequency and heavy-hitter queries by varying sketch size from 4KB to 4MB.}
    \label{fig-sketch-size}
    \Description{experiments}
\end{figure}

\subsection{Space and Time Efficiency}
\label{sec-exp-space-time}

Figure~\ref{fig-sketch-size} illustrates the performance of different sketches for frequency and heavy-hitter queries when their sizes range from 4KB to 4MB and $\varepsilon = 1$.
We observe that for CMS-based methods, i.e., \textsc{U-Sketch}, \textsc{DPSW-Sketch}, and \textsc{Microscope-Sketch}, their query performance first decreases with increasing sketch sizes and then becomes steady when the sketch sizes are sufficiently large because larger sketches naturally lead to fewer hash collisions and more accurate counters.
On the contrary, MG sketch-based methods, i.e., \textsc{PCC-MG} and \textsc{BLMZ-Sketch}, generally perform better when their sketch sizes are smaller because larger sketches introduce much more noise while having little benefit.

\begin{figure}[t]
    \centering
    \includegraphics[width=.6\linewidth]{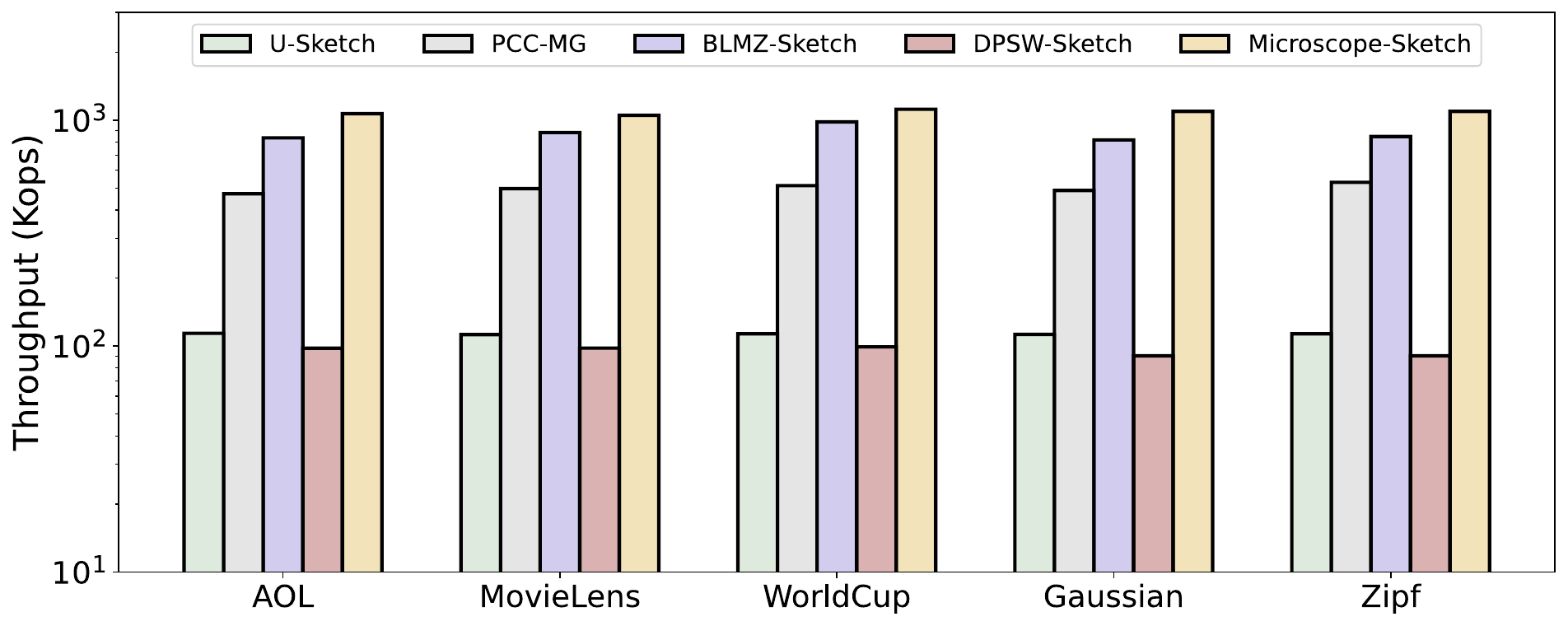}
    \vspace{-1em}
    \caption{Throughput of each sketch in the default setting.}
    \label{fig-throughput}
    \Description{experiments}
\end{figure}

Figure~\ref{fig-throughput} presents the throughputs of different sketches in the default setting (i.e., for $\varepsilon = 1$ and $w = 10^6$ with all the sketch parameters fine-tuned).
\textsc{DPSW-Sketch} has a throughput of about 100Kops (i.e., processing 100K items per second) across all datasets, which satisfies the requirement for real-world stream processing and is similar to that of \textsc{U-Sketch}.
However, \textsc{Microscope-Sketch}, \textsc{BLMZ-Sketch}, and \textsc{PCC-MG} have higher throughputs (500--1000Kops) than \textsc{DPSW-Sketch}.
For \textsc{Microscope-Sketch}, this is because it utilizes a more efficient structure without considering privacy.
For \textsc{BLMZ-Sketch} and \textsc{PCC-MG}, this can be attributed to the fact that MG sketches are much smaller than CMSs.

\section{Conclusion}
\label{sec-conclusion}

In this paper, we propose \textsc{DPSW-Sketch}, a novel sketch framework for frequency and heavy-hitter queries with event-level differential privacy (DP) in the sliding-window model.
Theoretically, \textsc{DPSW-Sketch} satisfies DP, provides results for both queries within bounded errors, and has sublinear time and space complexities w.r.t. the size of the sliding window.
These results improve upon those of existing private sliding-window sketches.
Experimental results also indicate that \textsc{DPSW-Sketch} achieves better trade-offs between utility and privacy than state-of-the-art baselines.

In future work, we would like to extend \textsc{DPSW-Sketch} to satisfy more rigorous notions of DP, such as user-level DP and local DP (LDP).
We are also interested in the design of private federated sketches for distributed data streams.

\bibliographystyle{ACM-Reference-Format}
\bibliography{ref}

\appendix

\section{Dataset Information}
\label{app-data}

\begin{figure}[t]
    \centering
    \includegraphics[width=\linewidth]{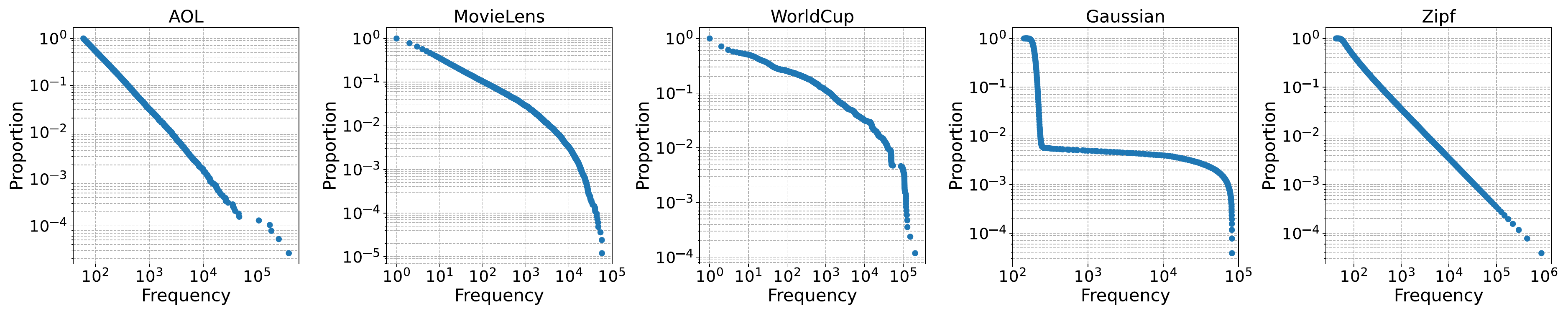}
    \vspace{-2em}
    \caption{Distribution of item frequencies in each dataset.}
    \label{fig-distribtuion}
    \Description{appendix}
\end{figure}

We describe the datasets used in our experiments and the pre-processing steps applied to them below.
\begin{itemize}
    \item \textbf{AOL}\footnote{\url{https://www.cim.mcgill.ca/~dudek/206/Logs/AOL-user-ct-collection/}} comprises 36,389,567 web queries over three months in the AOL website.
    Each record has several fields such as \texttt{AnonID}, \texttt{Query}, \texttt{QueryTime}, \texttt{ItemRank}, and \texttt{ClickURL}.
    We pre-processed the dataset as follows: (1) removing all duplicate and incomplete records; (2) sorting the records by \texttt{QueryTime}; (3) acquiring keywords from the \texttt{Query} field of each record after stopword removal and lemmatization; (4) filtering out typos and meaningless keywords. After all pre-processing steps, the dataset contains 10,710,880 items with 38,411 unique query keywords.
    We aim to keep track of the number of times each keyword is queried and the top trending keywords in the latest $w$ queries.
    \item \textbf{MovieLens}\footnote{\url{https://grouplens.org/datasets/movielens/latest/}} consists of 25,000,095 ratings on 62,423 movies contributed by 162,541 users from January 1995 to November 2019.
    Each record contains fields such as \texttt{userId}, \texttt{movieId}, \texttt{rating}, and \texttt{timestamp}.
    The original data set has been sorted chronologically without missing fields.
    We extract the \texttt{movieId} field for statistics to track the number of times each movie is rated and the most popular movies in the latest $w$ user interactions.
    \item \textbf{WorldCup}\footnote{\url{https://ita.ee.lbl.gov/html/contrib/WorldCup.html}} encompasses all 1,352,804,107 requests made to the 1998 FIFA World Cup website from April 30, 1998 to July 26, 1998.
    Each request has the following fields: \texttt{timestamp}, \texttt{clientID}, \texttt{objectID}, \texttt{size}, \texttt{method}, \texttt{status}, \texttt{type}, and \texttt{server}. Pre-processing steps include (1) removing incomplete and duplicate requests and (2) sorting the requests by \texttt{timestamp}.
    After pre-processing, the dataset contains 1,267,670,951 requests for 82,592 distinct URLs.
    We aim to track the number of times each URL is requested and the top trending URLs in the latest $w$ requests.
    \item \textbf{Gaussian} is a synthetic dataset consisting of 10,000,000 items with a domain size of 25,600, in which 95\% items are drawn from a Gaussian distribution with mean $\mu = 50$ and standard deviation $\sigma = 25$ (negative numbers are ignored) and rounded to the nearest positive integers, and the remaining 5\% items are drawn from a discrete uniform distribution in the range $[1, 25600]$.
    \item \textbf{Zipf} is also a synthetic dataset consisting of 10,000,000 items with a domain size of 25,600, in which 95\% items are drawn from a Zipf distribution with the skewness parameter $1$ and the remaining 5\% items are drawn from a discrete uniform distribution in the range $[1, 25600]$.
\end{itemize}
We illustrate the distribution of item frequencies in the above five datasets in Figure~\ref{fig-distribtuion}, where each point $(x, y)$ in a plot denotes that the ratio of the number of items with a frequency of at least $x$ and the number of items in the domain is $y$.

\section{Baseline Implementation}
\label{app-impl}

In the following, we provide some implementation details of different sketches in the experiments.
\begin{itemize}
    \item \textsc{PCC-MG} \cite{ChanLSX12}: There is no publicly available implementation of \textsc{PCC-MG}. Thus, we follow the specifications in \cite{ChanLSX12} to implement \textsc{PCC-MG}.
    Given a parameter $\lambda \in (0, 1)$, it builds a binary tree with $l = \lceil \ln \frac{4}{\lambda} \rceil$ levels, each of which consists of equal-length and disjoint blocks. For each block at the $i^{th}$ level, a (private) Misra-Gries sketch is constructed with the privacy parameter $\varepsilon_i = \frac{\varepsilon}{2^{l-i+1}}$ and the sketch parameter $\lambda_i = \frac{1}{2^i (l+1)}$.
    In each experiment, we test different values of $\lambda$ and report the best result achieved by \textsc{PCC-MG} among all values of $\lambda$.
    \item \textsc{U-Sketch} \cite{Upadhyay19}: To the best of our knowledge, \textsc{U-Sketch} has not been implemented for experimentation.
    We first tried to reproduce \textsc{U-Sketch} following the specifications in \cite{Upadhyay19}.
    However, this leads to large sketch sizes (typically several times the size of the sliding window), which is deemed unacceptable for stream processing.
    Therefore, we implement \textsc{U-Sketch} based on the general idea of \cite{Upadhyay19} but modify the parameter setting.
    To reduce space usage, we divide the sliding window into $100$ blocks and limit the number of sketches built in each block to $2$.
    Other parameters are fine-tuned for the best performance in each experiment.
    \item \textsc{BLMZ-Sketch} \cite{abs-2302-11081}: To the best of our knowledge, \textsc{BLMZ-Sketch} has also not been implemented for experimentation.
    We found that the specifications in \cite{abs-2302-11081} imply an excessively small block size (even less than $1$ in some cases).
    Therefore, similar to \textsc{U-Sketch}, we implement \textsc{BLMZ-Sketch} based on the general idea of \cite{abs-2302-11081} but modify the parameter setting.
    In particular, we fix the number $l$ of levels to $4$ and the size of each block to $\frac{w}{16}$.
    Other parameters are determined accordingly, and, in each experiment, we also fine-tune them and report the best result among all tested settings.
    \item \textsc{Microscope-Sketch} \cite{0001JDZCH0U023}: We use the code published by the authors on GitHub\footnote{\url{https://github.com/MicroscopeSketch/MicroscopeSketch}}.
    We test the variants of \textsc{Microscope-Sketch} using CMS, CU-Sketch, and CountSketch and find that the CMS version shows the best performance in most cases.
    Therefore, we only employ its CMS version in our experiments. Finally, in each experiment, we also report the best result among all parameter settings.
\end{itemize}

\section{Parameter Sensitivity}
\label{app-param}

\begin{figure}[t]
    \centering
    \includegraphics[width=\linewidth]{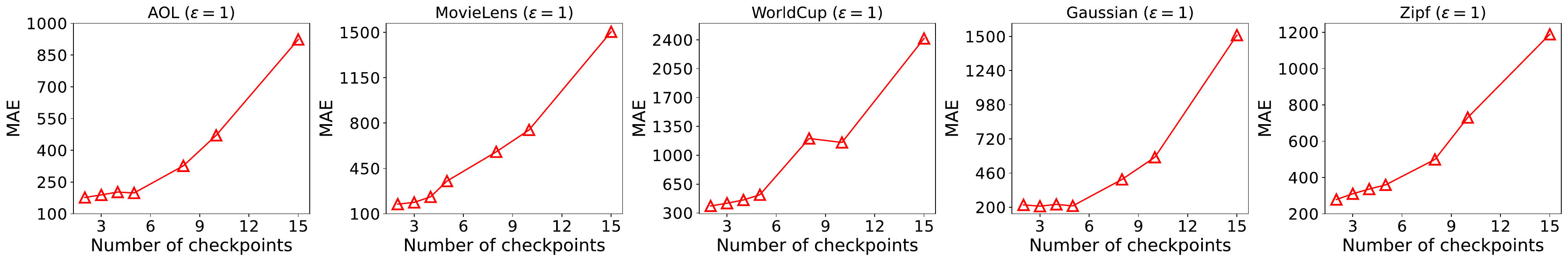}
    \includegraphics[width=\linewidth]{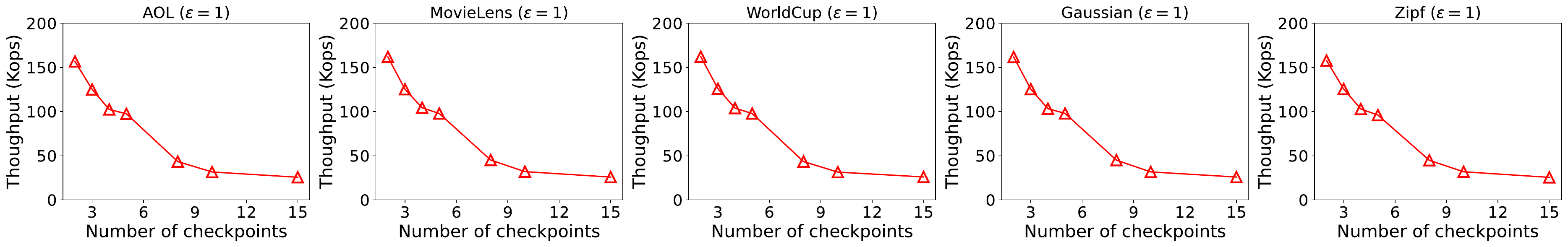}
    \vspace{-2em}
    \caption{Effect of the number of checkpoints in each smooth histogram on the performance of \textsc{DPSW-Sketch}.}
    \label{fig-tuning-checkpoint}
    \Description{appendix}
\end{figure}

Figure~\ref{fig-tuning-checkpoint} illustrates how the number of checkpoints in each smooth histogram affects the MAE for frequency estimation and throughput of \textsc{DPSW-Sketch}.
By setting the privacy parameter $\varepsilon$ to $1$ and the length of each sub-stream to $0.1 w$ and limiting the total sketch size to at most 100KB, we adjust the value of $\alpha$ from $0.99$ to $0.1$ so that $|I|$ is varied over $\{2, 3, 4, 5, 8, 10, 15\}$.
Our results indicate a significant growth in MAE across all datasets as the number of checkpoints increases.
This is because more checkpoints lead to fewer privacy budgets assigned to all PCMSs.
Accordingly, larger Gaussian noise should be added to each counter, which naturally causes higher MAEs.
Meanwhile, the throughput of \textsc{DPSW-Sketch} drops nearly linearly with the number of checkpoints since each item should be added to at most $2 |I| - 1$ PCMSs.
Therefore, we set the number of checkpoints $|I|$ to $3$ in the remaining experiments.

\begin{figure}[t]
    \centering
    \includegraphics[width=\linewidth]{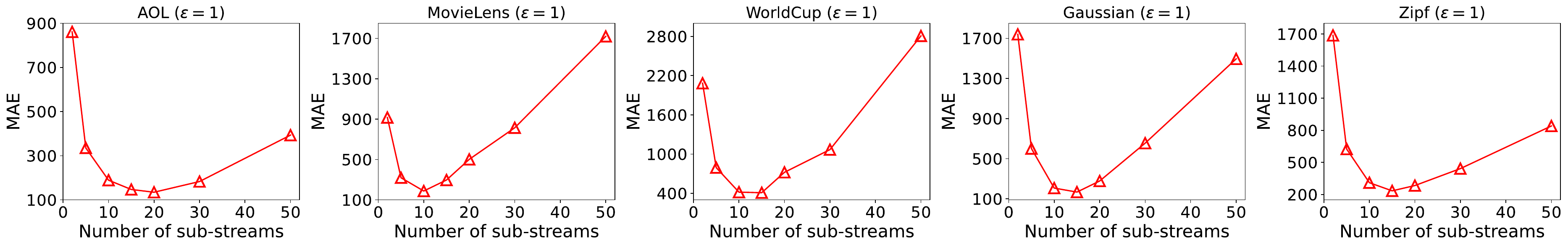}
    \includegraphics[width=\linewidth]{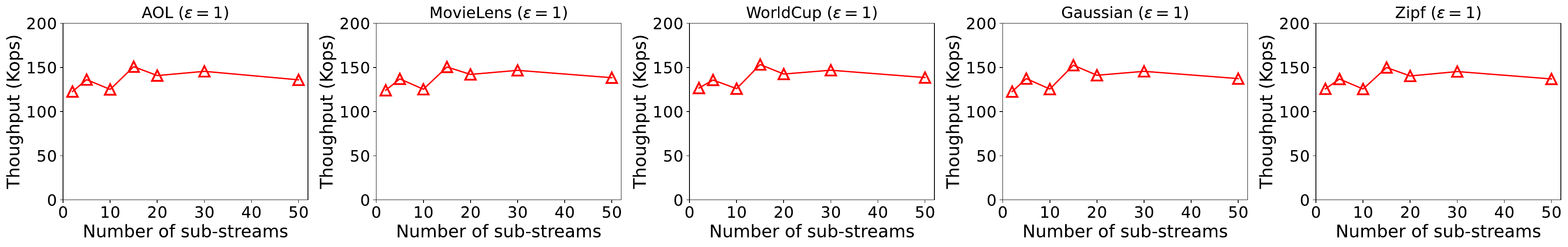}
    \vspace{-2em}
    \caption{Effect of the number of sub-streams in each window on the performance of \textsc{DPSW-Sketch}.}
    \label{fig-tuning-sub-num}
    \Description{appendix}
\end{figure}

Figure~\ref{fig-tuning-sub-num} illustrates the impact of the number of sub-streams in each window on the MAE of frequency estimation and throughput of \textsc{DPSW-Sketch}.
We fix $\varepsilon = 1$ and $|I| = 3$ and limit the total sketch size to at most 100KB.
We then vary the value of $\beta$ to adjust the length of each sub-stream from $0.02 w$ to $0.5 w$.
Consequently, the number of sub-streams in each window is set to $\{2, 5, 10, 15, 20, 30, 50\}$.
We find that, as the number of sub-streams increases, the MAE first decreases, reaching its lowest point around $10$ sub-streams per window, and then increases.
On the one hand, when the length of a sub-stream is too large, the high MAE is caused by the misalignment between the current window and the sketch used for query processing.
On the other hand, when the length of a sub-stream is too small, the limitation on the total sketch size requires that the size of each PCMS is smaller, leading to more hash collisions and a higher MAE.
The throughput is almost not affected by the number of sub-streams, as \textsc{DPSW-Sketch} only updates the PCMSs built in the latest sub-stream.
Therefore, we set the length of each sub-stream to $0.1 w$ in the remaining experiments.

\begin{figure}[t]
    \centering
    \includegraphics[width=\linewidth]{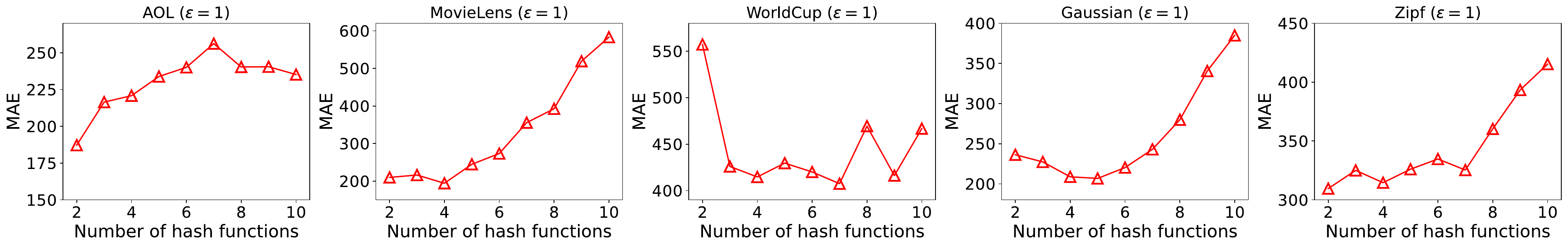}
    \includegraphics[width=\linewidth]{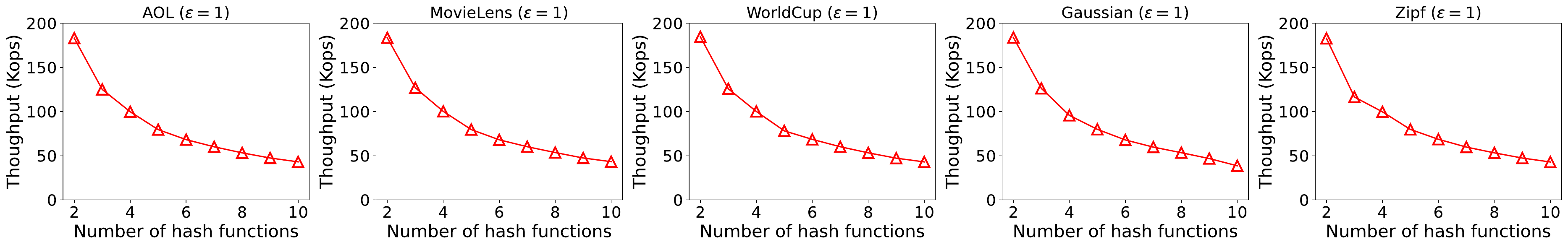}
    \vspace{-2em}
    \caption{Effect of the number of hash functions in each PCMS on the performance of \textsc{DPSW-Sketch}.}
    \label{fig-tuning-a}
    \Description{appendix}
\end{figure}

Figure~\ref{fig-tuning-a} shows how the values of $a$ and $b$ for each PCMS affect the MAE for frequency estimation and throughput of \textsc{DPSW-Sketch}.
We also fix $\varepsilon = 1$, $|I| = 3$, and the length of each sub-stream to $0.1 w$.
Then, by limiting the total sketch size to 100KB, we vary the number $a$ of hash functions in each PCMS from $2$ to $10$ and, accordingly, the range of each hash function from $5,000$ to $1,000$.
The results show that no single value of $a$ guarantees good performance in all datasets because each dataset has a different frequency distribution of items and domain size.
The throughput decreases almost linearly as the number of hash functions increases as each PCMS updates $a$ counters per item.
Since the MAE does not decrease when $a > 5$ in most datasets, we only test $a = 2, 3, 4, 5$ for high throughput and present the best result among them in each experiment.

\begin{figure}[t]
    \centering
    \includegraphics[width=\linewidth]{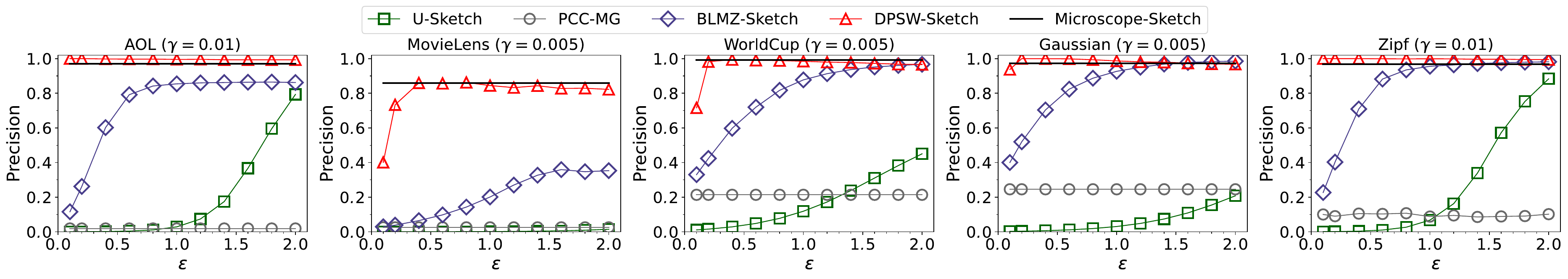}
    \includegraphics[width=\linewidth]{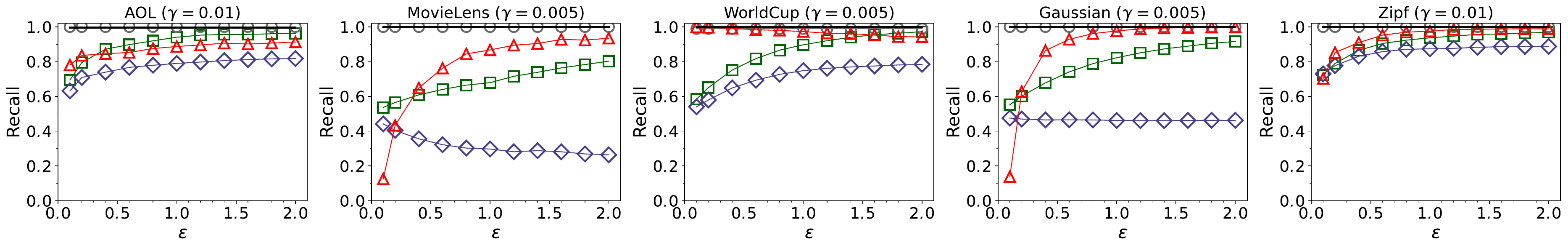}
    \includegraphics[width=\linewidth]{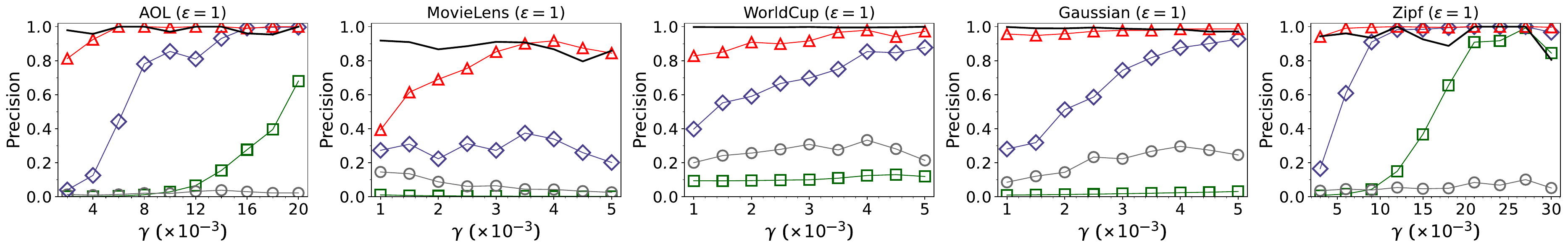}
    \includegraphics[width=\linewidth]{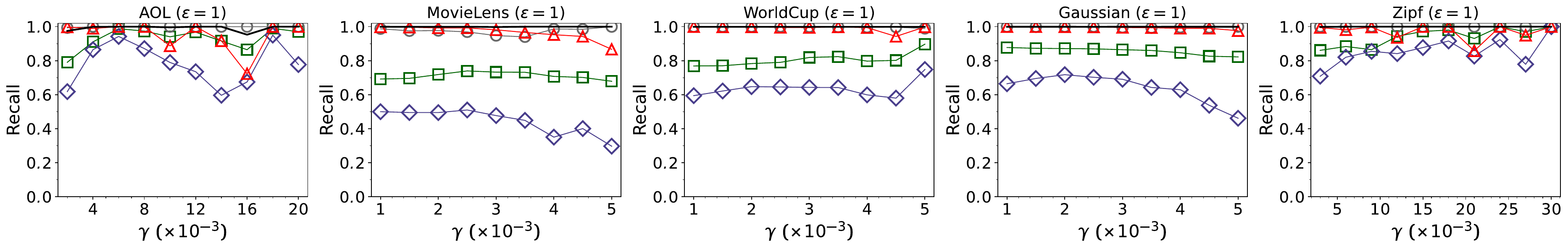}
    \vspace{-2em}
    \caption{Precision and recall scores for heavy-hitter identification by varying privacy parameter $\varepsilon$ and threshold $\gamma$.}
    \label{fig-pre-rec}
    \Description{appendix}
\end{figure}

\section{Supplemental Experiments}

\paragraph{Precision and Recall for Heavy-Hitter Identification}
To complement the experimental results in Section~\ref{sec-exp-hh}, we provide the precision and recall scores of different sketches for heavy-hitter identification by varying the privacy parameter $\varepsilon$ and the threshold $\gamma$ in Figure~\ref{fig-pre-rec} following the same settings as used in Figure~\ref{fig-hh}.
These results further verify the outstanding performance of \textsc{DPSW-Sketch} for heavy-hitter identification, as it simultaneously achieves higher precision and recall scores than \textsc{BLMZ-Sketch} and \textsc{U-Sketch}.
We also confirm that \textsc{PCC-MG} does not achieve high F1-scores due to extremely low precision scores, although it always has a recall score of $1$.
Finally, all methods have relatively low performance for heavy-hitter identification in the MovieLens dataset.
This is due to \emph{concept drift}: as new trending movies always emerge over time, the set of heavy hitters in the MovieLens dataset continuously evolves.
For comparison, item frequencies are almost independent of time in synthetic datasets; although trending keywords and URLs also change over time in the AOL and WorldCup datasets, they are quite stable relative to the length of the sliding window.
Since none of the compared methods explicitly considers the issue of concept drift, they cannot achieve as high performance in the MovieLens dataset as in other datasets.
We leave the problems of detecting changes in data distribution and taking into account concept drift in heavy-hitter identification with privacy concerns for future work.

\begin{figure}[t]
    \centering
    \includegraphics[width=\linewidth]{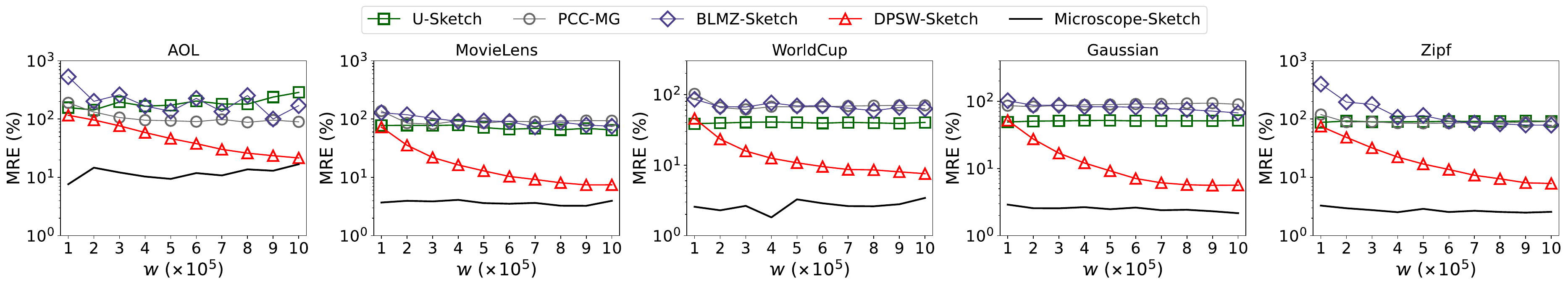}
    \includegraphics[width=\linewidth]{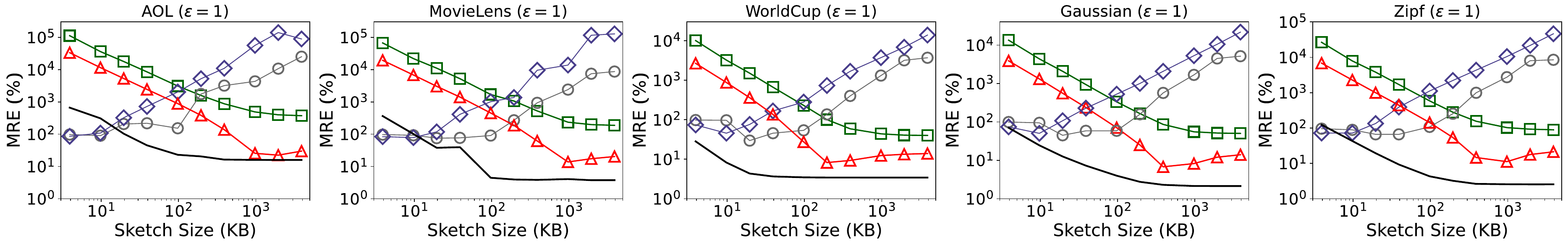}
    \vspace{-2em}
    \caption{MRE for frequency estimation by varying window size and sketch size.}
    \label{fig-sketchsize-mre-f1}
    \Description{appendix}
\end{figure}

\paragraph{MRE with Varying Window Size and Sketch Size}
We present the MRE of each sketch for frequency estimation by varying the window size and the sketch size in Figure~\ref{fig-sketchsize-mre-f1}, following the same setting as used in Figures~\ref{fig-window} and~\ref{fig-sketch-size}.
We observe similar trends as described in Sections~\ref{sec-exp-ws} and~\ref{sec-exp-space-time}.

\begin{figure}[t]
    \centering
    \includegraphics[width=.8\linewidth]{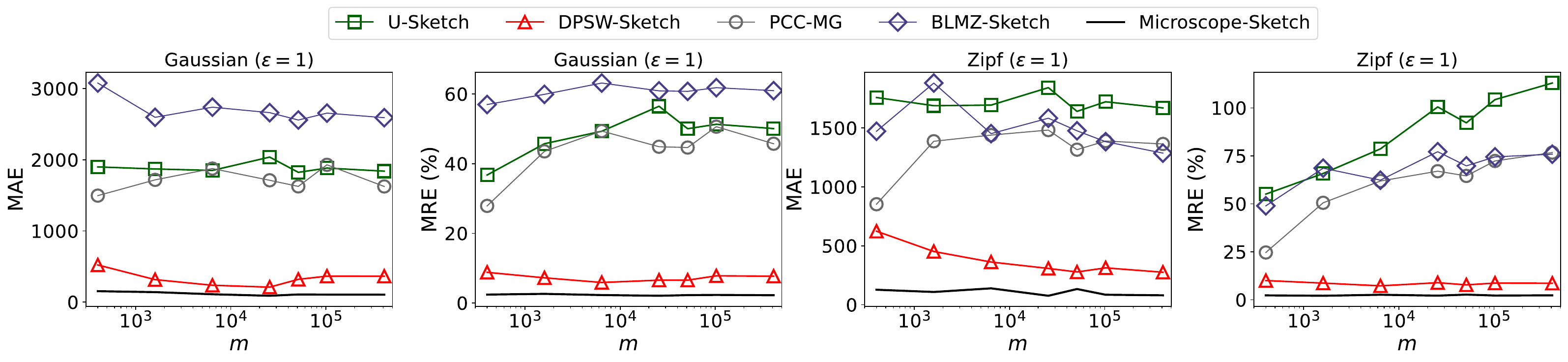}
    \vspace{-1em}
    \caption{MAE and MRE for frequency estimation by varying domain size $m$.}
    \label{fig-domain}
    \Description{appendix}
\end{figure}

\paragraph{Effect of Domain Size}
Figure~\ref{fig-domain} illustrates the impact of the domain size $m$ on the MAE and MRE for frequency estimation in the two synthetic datasets.
We generate Gaussian and Zipf datasets of 10M items with domain sizes ranging from $400$ to $409,600$ following the procedure in Appendix~\ref{app-data}.
The results indicate that the MAE for all methods remains relatively stable across different domain sizes.
\textsc{DPSW-Sketch} and \textsc{Microscope-Sketch} exhibit stability in MAE and MRE for different domain sizes, with \textsc{DPSW-Sketch} consistently outperforming other differentially private baselines and closely matching the performance of \textsc{Microscope-Sketch}.
Generally, \textsc{DPSW-Sketch} shows strong scalability against the dimensionality of the datasets.

\end{document}